\documentclass[sigconf,prologue,dvipsnames]{acmart} %
\usepackage{url}
\usepackage{latexsym}
\usepackage{amsmath}
\usepackage{graphicx}
\usepackage{subcaption}
\usepackage{booktabs}
\usepackage{xcolor}
\usepackage{multirow}
\usepackage[inline]{enumitem}
\usepackage{array}
\usepackage{xspace}
\usepackage{textgreek}

\definecolor{dark-gray}{gray}{0.33} 
\newcommand{\cut}[1]{}
\newcommand{\deem}[1]{\textcolor{dark-gray}{\small#1}}

\newcolumntype{P}[1]{>{\centering\arraybackslash}p{#1}}

\newif\ifshowcomments
\showcommentstrue
\newcommand\todo{\textcolor{red}}
\newcommand{\cd}[1]{{\textcolor{blue}{#1}}}
\newcommand{\jpc}[1]{{\textcolor{magenta}{#1}}}
\newcommand{\justin}[1]{{\textcolor{purple}{#1}}}
\ifshowcomments
\else
\renewcommand\todo[1]{}
\renewcommand{\cd}[1]{}
\renewcommand{\jpc}[1]{}
\renewcommand{\justin}[1]{}
\fi

\newcommand{\xhdr}[1]{{\noindent\bfseries #1.}}

\newcommand{\intenttype}{goal\xspace}

\copyrightyear{2020}
\acmYear{2020}
\setcopyright{iw3c2w3}
\acmConference[WWW '20]{Proceedings of The Web Conference 2020}{April 20--24, 2020}{Taipei, Taiwan}
\acmBooktitle{Proceedings of The Web Conference 2020 (WWW '20), April 20--24, 2020, Taipei, Taiwan}
\acmPrice{}
\acmDOI{10.1145/3366423.3380273}
\acmISBN{978-1-4503-7023-3/20/04}

\title[Don't Let Me Be Misunderstood]{Don't Let Me Be Misunderstood:\\Comparing Intentions and Perceptions in Online Discussions}

\author{Jonathan P. Chang}
\affiliation{%
  \institution{Cornell University}
}
\email{jpc362@cornell.edu}

\author{Justin Cheng}
\affiliation{%
  \institution{Facebook}
}
\email{jcheng@fb.com}
\author{\mbox{Cristian~Danescu-Niculescu-Mizil}}
\affiliation{%
  \institution{Cornell University}
}
\email{cristian@cs.cornell.edu}

\begin{document}

\begin{abstract}
Discourse involves two perspectives: a person's intention in making an utterance and others' perception of that utterance.
The misalignment between these perspectives can lead to undesirable outcomes, such as misunderstandings, low productivity and even overt strife.
In this work, we present a computational framework for exploring and comparing both perspectives in 
 online public discussions.

We combine logged data about public comments on Facebook with a survey of over 16,000 people about their intentions in writing these comments or about their perceptions of comments that others had written.
Unlike previous studies of online discussions that have largely relied on third-party labels to quantify properties such as sentiment and subjectivity, our approach also directly captures what the speakers actually intended when writing their comments.
In particular, our analysis focuses on judgments of whether a comment is stating a fact or an opinion, since these concepts were shown to be often confused.

We show that intentions and perceptions diverge in consequential ways.
People are more likely to perceive opinions than to intend them, and linguistic cues that signal how an utterance is intended can differ from those that signal how it will be perceived.
Further, this misalignment between intentions and perceptions can be linked to the future health of a conversation:
when a comment whose author intended to share a fact is misperceived as sharing an opinion, the subsequent conversation is more likely to derail into uncivil behavior than when the comment is perceived as intended.
Altogether, these findings may inform the design of discussion platforms that better promote positive interactions.

\end{abstract}

\maketitle

\section{Introduction}
\label{sec:intro}
\begin{quote}
\small
\vspace{-0.01cm}
``I'm just a soul whose intentions are good...''
\end{quote}
\begin{flushright}
\small
-- Nina Simone, ``Don't Let Me Be Misunderstood''
\end{flushright}

Conversations, both online and offline, fundamentally involve two perspectives: a speaker's \emph{intention}---that is, the
 goals they seek to achieve through their utterance---and others' \emph{perception} of the speaker's words~\cite{grosz_attention_1986}.
When the intentions and perceptions of participants in a conversation are misaligned \cite{clark_using_1996,tannen_conversational_2005}, undesirable outcomes ranging from low productivity to overt strife can occur~\cite{mckee_your_2002,tannen_indirectness_2000}.

One important type of misalignment occurs when people confuse \emph{facts}\footnote{Following prior literature, we consider facts to be statements that could in principle be conclusively proven or disproven based on evidence regardless of their veracity~\cite{mitchell_distinguishing_2018,corral-verdugo_effect_1993}.} and \emph{opinions} \cite{rabinowitz_distinguishing_2013}.
A Pew survey on online news consumption found that 65\% of Americans mistakenly perceived opinions extracted from online news as facts, while 75\% took facts to be opinions \cite{mitchell_distinguishing_2018}.
In this work, we investigate how this type of misalignment plays out in online public discussions where people engage with each other rather than passively consuming content.
How often and under what circumstances does a speaker's intended statement of a fact get misperceived as an opinion?
How does such misalignment tie into the quality of online discourse?

Answering such questions requires ground truth data both on what the speaker's intention was in crafting an utterance and on how that utterance was perceived by others.
While perceptions can be approximated through third-party annotation~\cite{wiebe_learning_2004,lex_objectivity_2010,regmi_what_2015}, only the speakers themselves know what their actual intentions were.
To obtain ground truth about both intention and perception, we surveyed
over 16,000
people about their intention in writing public comments on Facebook or about how they perceived comments to which they had replied.
Combining these surveys with data about these conversations then allows us to compare how facts and opinions are intended and perceived in different contexts.
We start with a high-level approach exploring differences in the distributions of intentions and perceptions.
 We find, for example, that in online discussions people perceive opinions at a higher rate than they are intended.
Next, we investigate this apparent incongruency at a linguistic level.
While linguistic cues developed to capture subjectivity can generally distinguish between facts and opinions, salient differences arise when considering how utterances are intended rather than how they are perceived.
For instance, the explicit use of factual language (e.g., "In fact,\ ...") signals that the speaker intended to make a factual claim, but not that 
others
 will perceive it as such.
Starting from these insights,
we 
assess
 the extent to which third-party perception labels (as used in prior work on subjectivity detection) are interchangeable with author-sourced intention labels when predicting intentions from text.
Finally, we examine how differences between intentions and perceptions relate to the outcome and quality of online public discussions. The trajectory of a conversation likely depends on its intended starting point as well as on 
how others perceive it.
For example, past work on Wikipedia discussions found that conversations starting with factual checks are more likely to turn uncivil than those appearing to share or seek opinions~\cite{zhang_conversations_2018},
 and
  qualitative studies have suggested that the alignment between the intentions and perceptions of participants in a conversation is key to keeping it on track~\cite{tannen_indirectness_2000,clark_using_1996}.

We find that both the intended and perceived \emph{goals} of the initial comment are indicative---in potentially different ways---of a conversation's future trajectory.
While certain intended goals (e.g., sharing an opinion) are more likely to lead to uncivil behavior in the subsequent discussion, even more significant is whether other participants in the conversation perceive those goals as they were intended.
For instance, when the 
initial commenter intends to share a fact but this goal is \emph{misperceived} by others, the conversation is more likely to 
turn uncivil
than when that initial intention is correctly perceived.

Taken together, these findings have potential implications for designing public discussion platforms that can better promote healthier conversations.
They show that when assessing the likely trajectory of a conversation, both a speaker's intention and how that intention might be (mis)perceived by others are important factors to consider.
More generally, by exposing differences between ground-truth labels coming from a first-person perspective and those coming from a (more easily accessible) third-person perspective, these results also signal potential biases that systems relying only on one of the two types of labels might incur.

To summarize, we:
\begin{itemize}
    \item Present a large-scale study comparing intentions and perceptions of facts and opinions in online conversations and their relation to conversational outcomes;
    \item Identify and compare linguistic cues that signal intentions and perceptions,
    showing that linguistic differences between the two translate to differences in 
    classifier behavior when training on one versus the other; and
    \item Show that conversations in which fact-sharing intentions are misperceived as opinion-sharing are more likely to turn uncivil later on.
\end{itemize}

\section{Ground Truth Intentions and Perceptions}
\label{sec:data}
Though studies of online discussions often 
use third-party annotation to identify properties of interest (e.g., opinions \cite{wiebe_annotating_2005}), this can only capture perceptions and not intentions, as only the original author of a comment knows with certainty 
what they intended.
As such, we instead surveyed comment authors directly to gather intention labels.
In this section, we describe our 
conversational data
and survey methodology, and 
give high-level descriptive statistics of the survey responses.
All data was de-identified and analyzed on Facebook's servers, and an internal research board reviewed the study design ethics and privacy practices prior to its start.

\subsection{Conversational data}

This work focuses on public discussions taking place in the comments sections of posts on Facebook Pages, which typically represent
brands, media outlets (including but not limited to news), public figures, or communities.
Anyone can view or join these discussions, so 
they offer a diverse sample of data for comparing intentions and perceptions of facts and opinions.

But as replies are rare on social media \cite{artzi_predicting_2012}, most comments are unlikely to be part of conversations.
As a heuristic for finding conversations in comment sections, we searched for a reciprocity pattern:
one person (the \emph{initiator}) makes a comment, a different person (the \emph{replier}) replies to the initiator's comment, and then the initiator follows up by either reacting to or replying to the replier.

We constructed our conversational dataset by finding comment threads on English-language Page posts that begin with this reciprocity pattern.
Data was collected from a 1.5 month window spanning mid-May through June 2019, resulting in approximately 22 million candidate conversations
taking place across 3 million posts on nearly 800,000 pages.

\subsection{Intentions and perceptions surveys}

\xhdr{Selecting survey participants}
Starting from this conversational dataset, we
created two survey participant pools: a pool of initiators who would receive a survey asking about their intentions, and a pool of repliers who would receive a survey asking about what they perceived to be the intention underlying the comment they replied to.
To minimize demand effects, we filtered out conversations where the initiator and replier were friends on Facebook.
We additionally filtered out any cases where at least one comment was no longer publicly viewable.
To ensure diversity in the participants and types of conversations we asked about in the surveys, we imposed a limit on how many participants could be selected from any given Page: up to 1\% of a Page's followers, capped at 10.

\begin{table*}
    \renewcommand*{\arraystretch}{1.10}
    \begin{tabular}{p{0.12\linewidth}p{0.41\linewidth}p{0.41\linewidth}}
        \toprule
        \textit{Question} & \textit{Initiator survey} & \textit{Replier survey} \\
        \midrule
        \multicolumn{3}{c}{\textbf{Goals}} \\
        \midrule
        Opinion sharing & When you started the interaction, were you trying to express an opinion? & Do you think the other person was trying to express an opinion? \\
        Opinion seeking & When you started the interaction, were you looking for other people's opinions? & Do you think the other person was looking for opinions? \\
                Fact sharing & When you started the interaction, were you trying to provide information (for example, sharing a fact)? & Do you think the other person was trying to provide information (for example, sharing a fact)? \\
        Fact seeking & When you started the interaction, were you looking for information? & Do you think the other person was looking for information? \\
        Humor & When you started the interaction, were you trying to make a joke? & Do you think the other person was joking? \\[-0.1em]
        & \deem{Not at all / Mostly not / Somewhat / Mostly / Definitely} & \deem{Not at all / Mostly not / Somewhat / Mostly / Definitely} \\
        \midrule
        \multicolumn{3}{c}{\textbf{Outcomes}} \\
        \midrule
        Time-worthiness & Looking back on this interaction, do you think it was worth your time? & Looking back on this interaction, do you think it was worth your time? \\[-0.1em]
        & \deem{Not at all / Mostly not / Somewhat / Mostly / Definitely} & \deem{Not at all / Mostly not / Somewhat / Mostly / Definitely} \\
        Understanding & How well do you feel your goals and intentions in this interaction were understood? & Overall, how difficult was it for you to guess the other person's goals and intentions? \\[-0.1em]
        & \deem{Not understood at all / Not very well understood / Somewhat understood / Mostly understood / Very well understood} & \deem{Not difficult at all / Not very difficult / Moderately difficult / Very difficult / Extremely difficult} \\
        \bottomrule
    \end{tabular}
    \caption{The initiator survey asked participants about their intentions with respect to a comment they had written, while the replier survey asked participants about their perceptions of a comment that they had replied to. Answers are shown in gray.} 
    \label{tab:surveys}
\end{table*}

\xhdr{Survey design}
The surveys for initiators and for repliers both asked about facts and opinions in the initiator's opening comment (i.e., the first comment of the reciprocal chain).
While in the context of monologic text---such as news articles and reviews---subjectivity mainly concerns the \emph{sharing} of facts versus opinions \cite{wiebe_annotating_2005,lex_objectivity_2010,yu_towards_2003},
in a conversational setting, participants can also 
\emph{seek} factual information or others' opinions \cite{morris_what_2010,quarteroni_simultaneous_2011}.
As such, both surveys
distinguished between sharing and seeking of opinions and facts.
Finally, prior work identifies humor as a prominent axis that is orthogonal to
opinions and facts \cite{moor_flaming_2010}, so we additionally 
included it in our survey for completeness.\footnote{In our findings, humor is nevertheless rare, being both intended and perceived in only about 10\% of cases, and so is excluded from most subsequent analyses.}
These considerations result in five \emph{\intenttype{s}}
that could be intended or perceived: (1) fact sharing, (2) fact seeking, (3) opinion sharing, (4) opinion seeking, and (5) humor.

The \emph{initiator survey} asked initiators to rate their opening comment along each \intenttype using a five-point Likert scale.
Analogously, the \emph{replier survey} asked repliers to rate their interpretation of the initiator's comment along each \intenttype.
Some subsequent analyses will simplify the responses by binarizing them: we will say that an initiator \emph{intended (or perceived) a \intenttype} if they responded with ``mostly'' or ``definitely'', and \emph{did not intend (or perceive) a \intenttype} otherwise.

\begin{figure}
    \centering
    \includegraphics[width=0.95\linewidth]{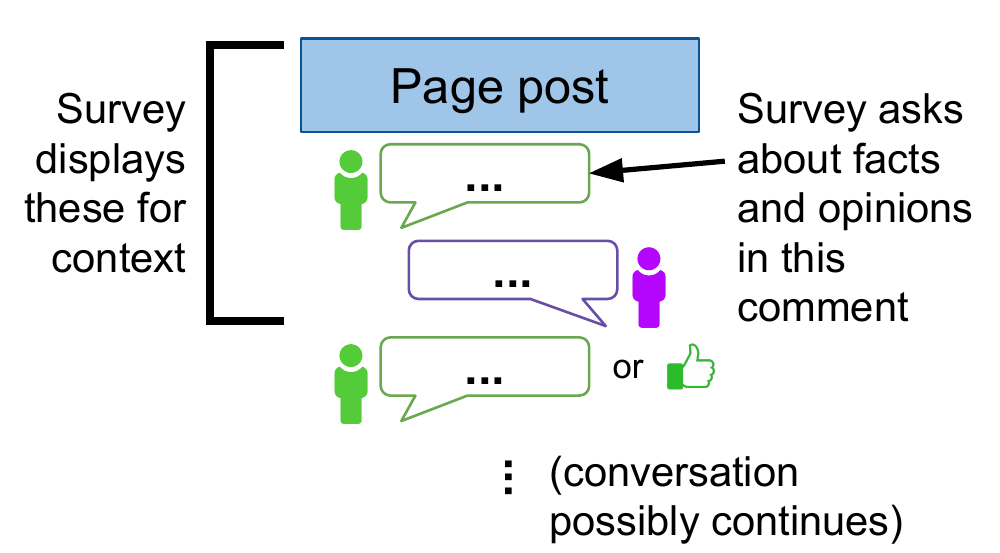}
    \caption{We surveyed on conversations containing \emph{reciprocity}: an \textcolor{ForestGreen}{initiator (green)} makes a comment, a \textcolor{violet}{replier (purple)} replies to the initiator, and the initiator follows up with another comment or a reaction. Surveys asked about facts and opinions in the initiator's opening comment, though for context the survey participant was additionally shown the reply and the Page post on which the exchange took place.}
    \label{fig:survey_convo}
\end{figure}

To better understand the relationship between intentions and conversational outcomes, both surveys also asked participants to rate the conversation along two axes: whether it was worth their time, and whether they felt understood (initiators) or found the other person easy to understand (repliers).
This results in a total of seven questions per survey (Table \ref{tab:surveys}).
Although both surveys asked about the initiator's opening comment, for context the survey participants were also shown the reply and the Page post on which the conversation took place (Figure \ref{fig:survey_convo}).
Low response rates made it infeasible to obtain paired initiator-replier responses (i.e., responses from both the initiator and replier on each conversation).
As a result, we ran the initiator and replier surveys on disjoint conversations.
While we conducted third-party annotation to study the relationship between intentions and perceptions in the same conversation (Section \ref{sec:eval}), exploring other ways to address this limitation would be valuable future work.

\xhdr{Running the survey}
Participants were recruited for both surveys via an ad on Facebook targeted at a random sample of English-speaking people, which ran for two weeks in early July 2019.
Each survey was opt-in, and participants could choose to stop at any time.
Other than the ad and survey, participants' Facebook experience was not altered or manipulated in any way.

\begin{figure*}
    \centering
    \begin{subfigure}[b]{0.24\textwidth}
        \includegraphics[width=\textwidth]{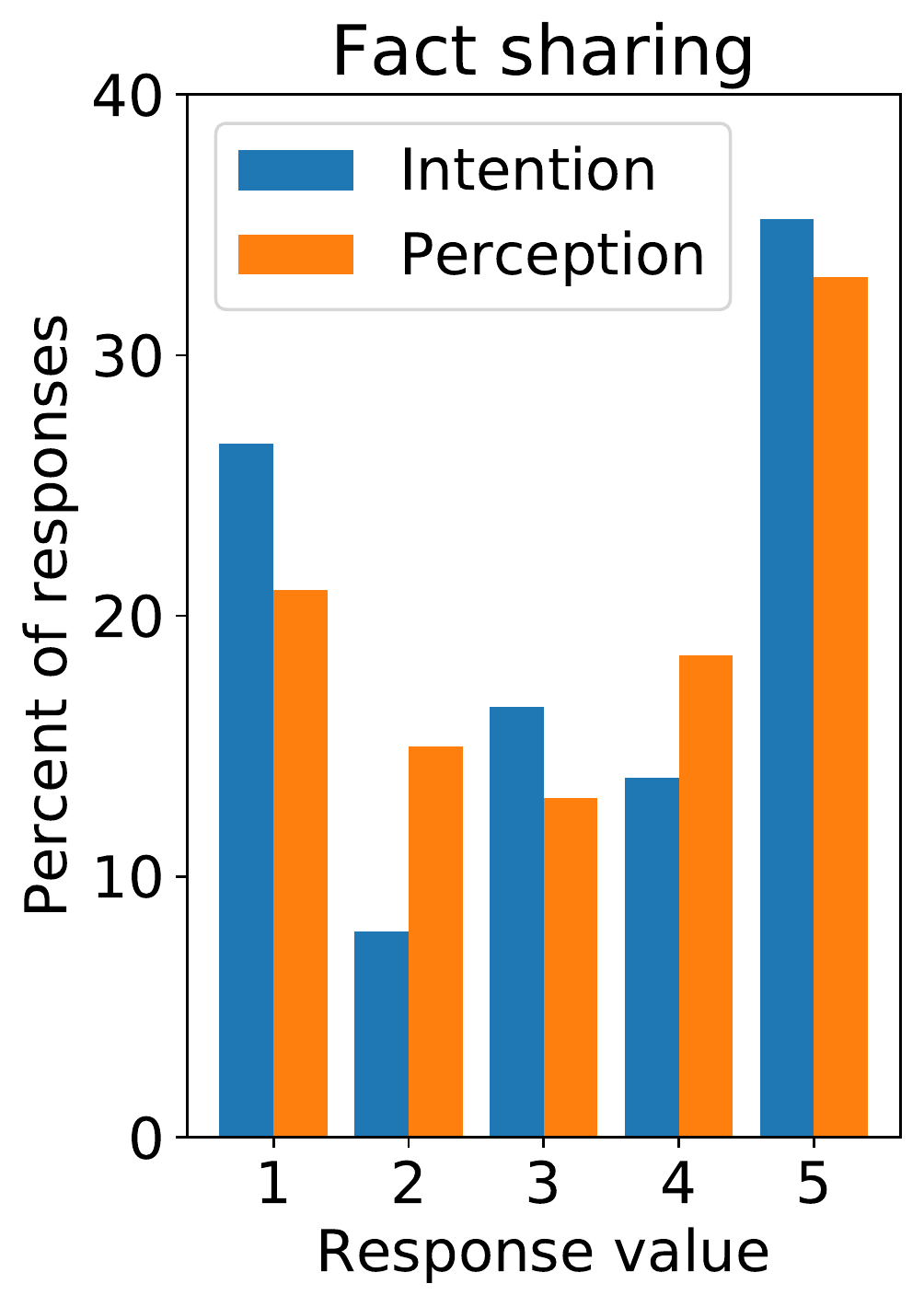}
        \caption{}
        \label{fig:info_giving_dist}
    \end{subfigure}
    \begin{subfigure}[b]{0.24\textwidth}
        \includegraphics[width=\textwidth]{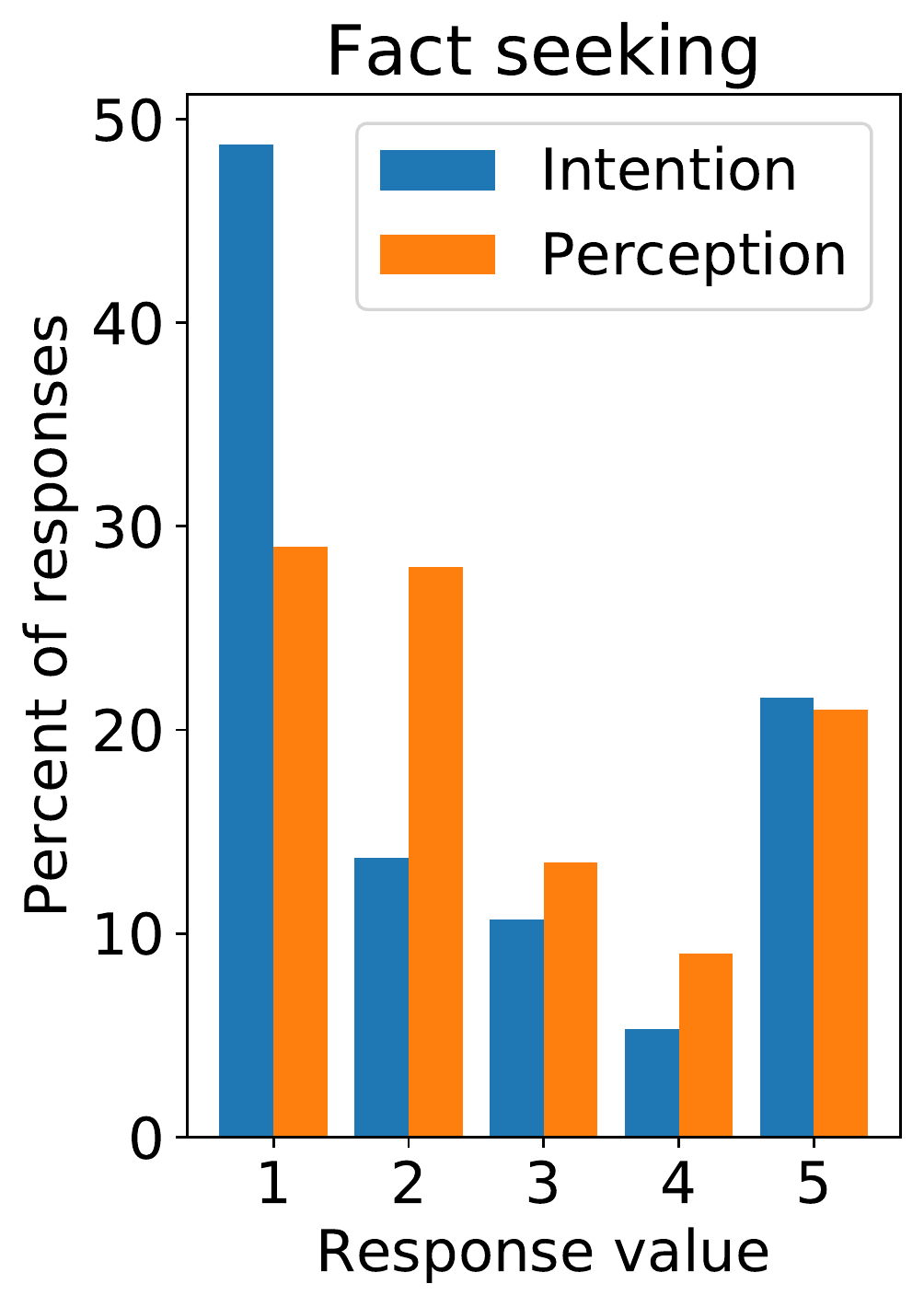}
        \caption{}
        \label{fig:info_seeking_dist}
    \end{subfigure}
    \begin{subfigure}[b]{0.24\textwidth}
        \includegraphics[width=\textwidth]{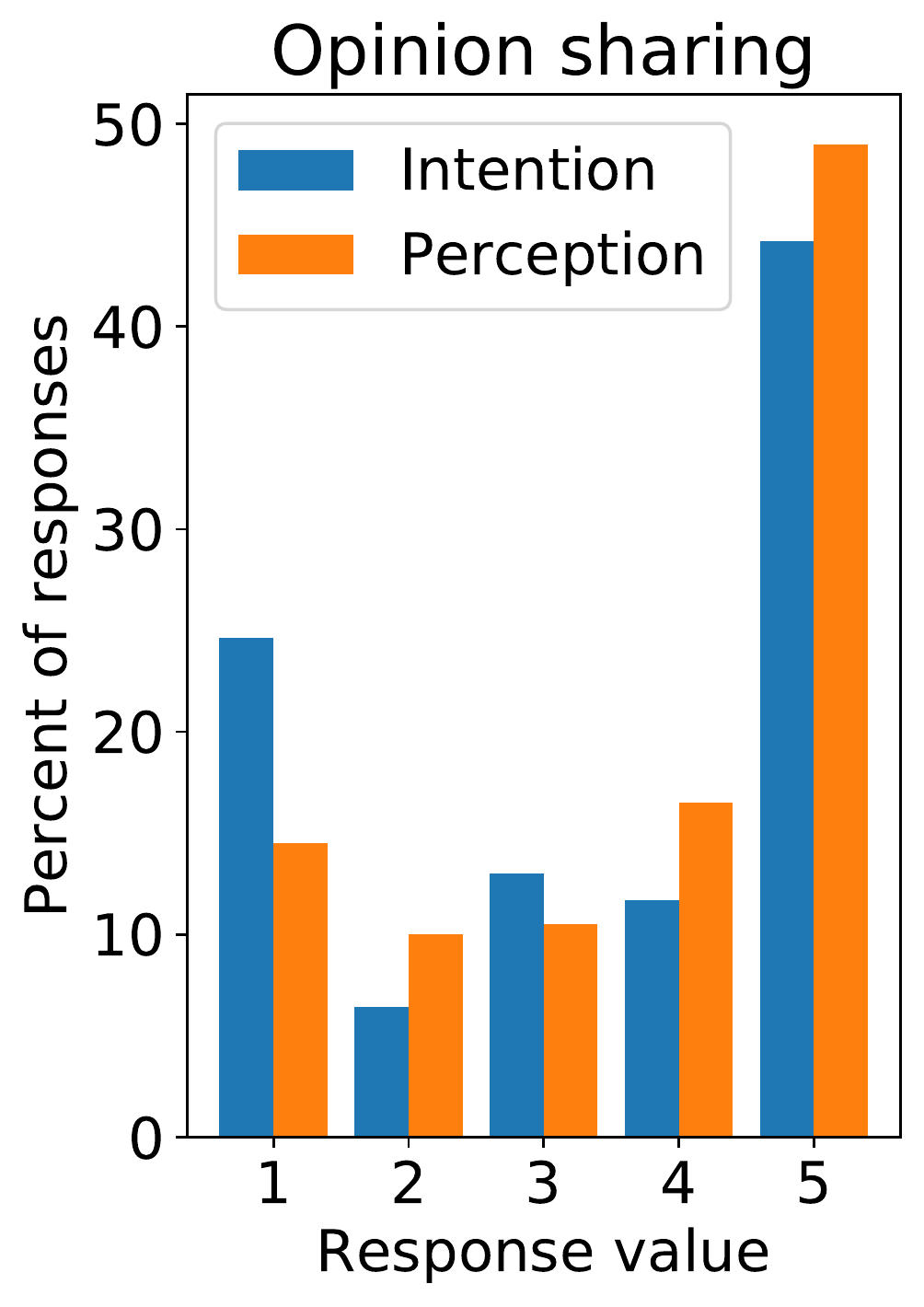}
        \caption{}
        \label{fig:opinion_giving_dist}
    \end{subfigure}
    \begin{subfigure}[b]{0.24\textwidth}
        \includegraphics[width=\textwidth]{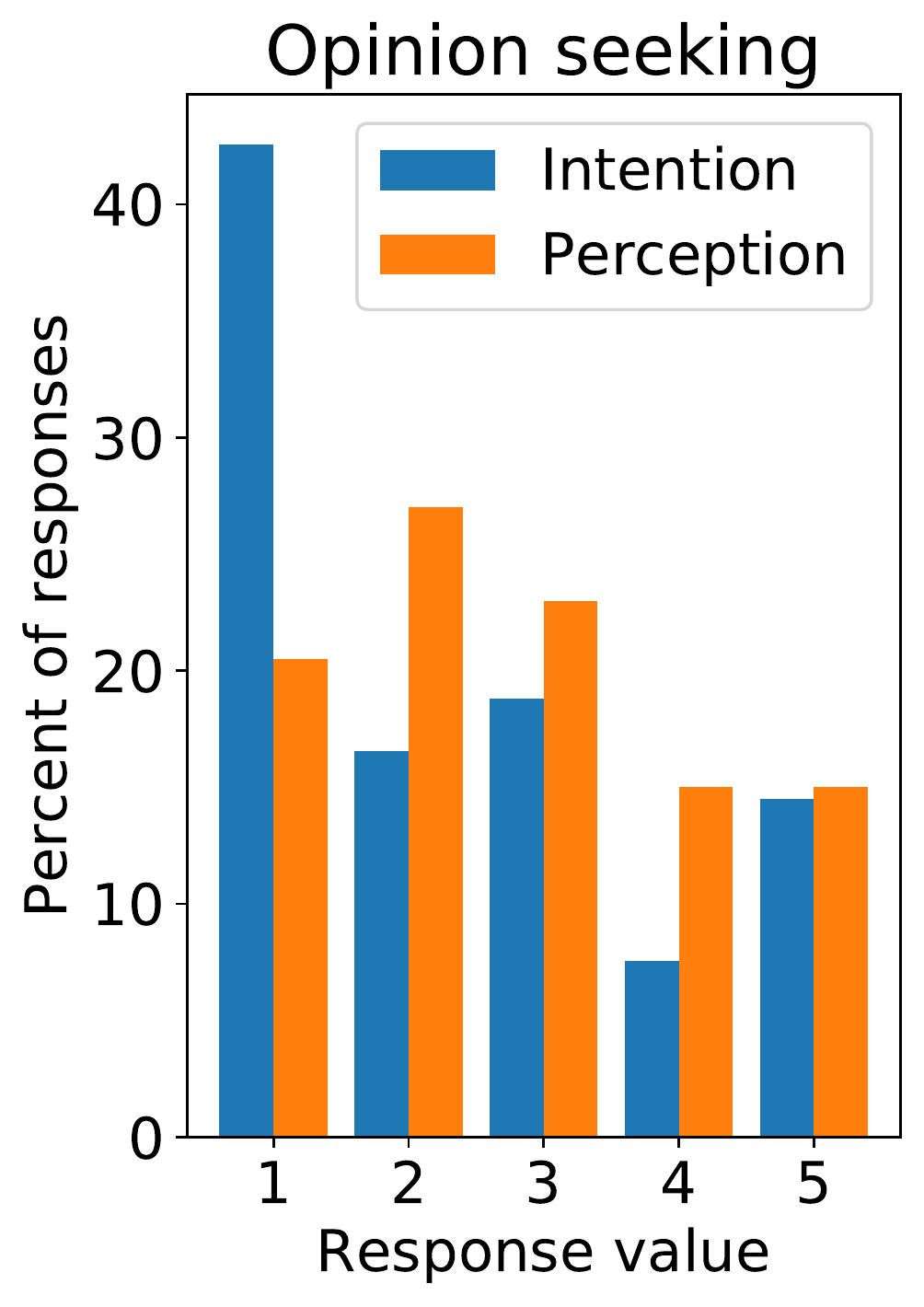}
        \caption{}
        \label{fig:opinion_seeking_dist}
    \end{subfigure}
    \caption{
    Distributional differences in how people intend their own comments and perceive others' comments. For example, people perceive an opinion sharing goal more often than it is intended (c). A response value of 1 corresponds to ``did not intend at all'', while a response value of 5 corresponds to ``definitely had this intent''.
    }
    \label{fig:intent_perception_dists}
\end{figure*}

\subsection{Participant statistics}
9,174 people completed the initiator survey, while 7,129 people completed the replier survey.
On average, participants were 5.6 years older and 1.6\% more likely to be female compared to the average Facebook Page commenter.
To test if the amount of time between when someone commented or replied and when they took the survey affected their responses, we calculated correlations with this time gap, finding that they are negligible ($r \le 0.02$, n.s.).

The lower response rate for the replier survey suggests that questions about one's own intentions are easier to answer than questions about perceiving others' intentions.
To test for a response bias, we examined demographic differences between the two surveys.
We find small but significant differences for age ($D = 0.03$, $p < 0.001$ using a K-S test) and gender ($\chi^2$ = 3.97, $p = 0.05$).
Though these differences are small, they may have an effect on results if responses vary significantly across demographics.
Thus, we next examine how age and gender may affect intentions and perceptions.

Older people are less likely to intend to seek facts (Spearman's R = -0.08, $p < 0.001$) or opinions (Spearman's R = -0.09, $p < 0.001$),
which may be partly explained by previous work showing that older people tend to prefer passive learning (e.g., through reading) over learning through direct interaction \cite{giambra_curiosity_1992}.
They are also more likely to perceive others as sharing facts (Spearman's R = 0.12, $p < 0.001$), echoing prior research that showed older people may be more inclined to treat a statement as factual \cite{guess_less_2019}.
We also find that men were less likely to intend to seek opinions from others (Mann-Whitney U = 10820738.5, $p < 0.001$).
To address these potentially confounding effects, in subsequent analyses we control for demographic differences (as well as descriptive properties of the Page, namely size and category) as appropriate.
We further note that although the demographic effects are statistically significant, demographic and Page features are nonetheless poor predictors of intention and perception,\footnote{$R^2 \le 0.04$ in regressions predicting intention or perception using gender, age, Page size (number of followers), and Page category (e.g., ``sports'').} 
and in practice we find that uncontrolled versions of each analysis yield similar results.

\section{Intentions versus perceptions}

\label{sec:differences}

To understand if there is a systematic misalignment between intentions and perceptions in the context of online public discussions, we
\begin{enumerate*}[label={(\alph*)}]
\item compare response frequencies among the initiator survey responses and replier survey responses (i.e., how often a \intenttype is actually intended versus how often it is perceived),
\item consider linguistic cues that are indicative of a \intenttype and explore whether these are different for
intended versus perceived \intenttype{s}, and
\item examine how intentions and perceptions may differ in their relationship to the trajectory of the conversations in which they are observed.
\end{enumerate*}

\subsection{Distributional differences}
If perceptions perfectly captured intentions, the overall distribution of responses for intentions and perceptions of each \intenttype would be nearly identical.
But if intentions and perceptions are misaligned, then we may observe systematic differences between the two response distributions.
As such, our first analysis compares response distributions of intention and perception for each \intenttype.
We controlled for demographic differences between surveys by reweighting the perception survey responses
to match the age and gender distribution of the intention survey via post-stratification \cite{valliant_poststratification_1993}. 
These distribution comparisons are visualized in Figure \ref{fig:intent_perception_dists}.
This data exposes
two types of distributional differences:
\begin{enumerate*}[label={(\alph*)}] %
\item systematic overestimation, in which perceivers judge a particular
\intenttype to occur more frequently than it is actually intended, and
\item uncertainty, in which perceivers are unsure of people's intentions and hence
tend to pick less definite response choices.
\end{enumerate*}

\xhdr{Systematic overestimation}
Systematic overestimation occurs when a \intenttype is perceived to occur more frequently than it is actually intended.
This can be formalized as the mean response for perception being significantly larger than the mean response for intention.
Under this definition, overestimation occurs for opinion sharing (mean perception response = 3.8, mean intention response = 3.4), corroborating prior work which found that people were more likely to misidentify factual statements as opinions than 
vice versa
\cite{mitchell_distinguishing_2018,rabinowitz_distinguishing_2013}.
Overestimation also occurs for fact seeking (2.7 vs 2.4) and opinion seeking (2.8 vs 2.4), but not fact sharing.
These differences are significant at $p < 0.001$ via Mann-Whitney U test.

\begin{figure*}
    \centering
    \begin{subfigure}[b]{0.48\textwidth}
        \includegraphics[width=\textwidth]{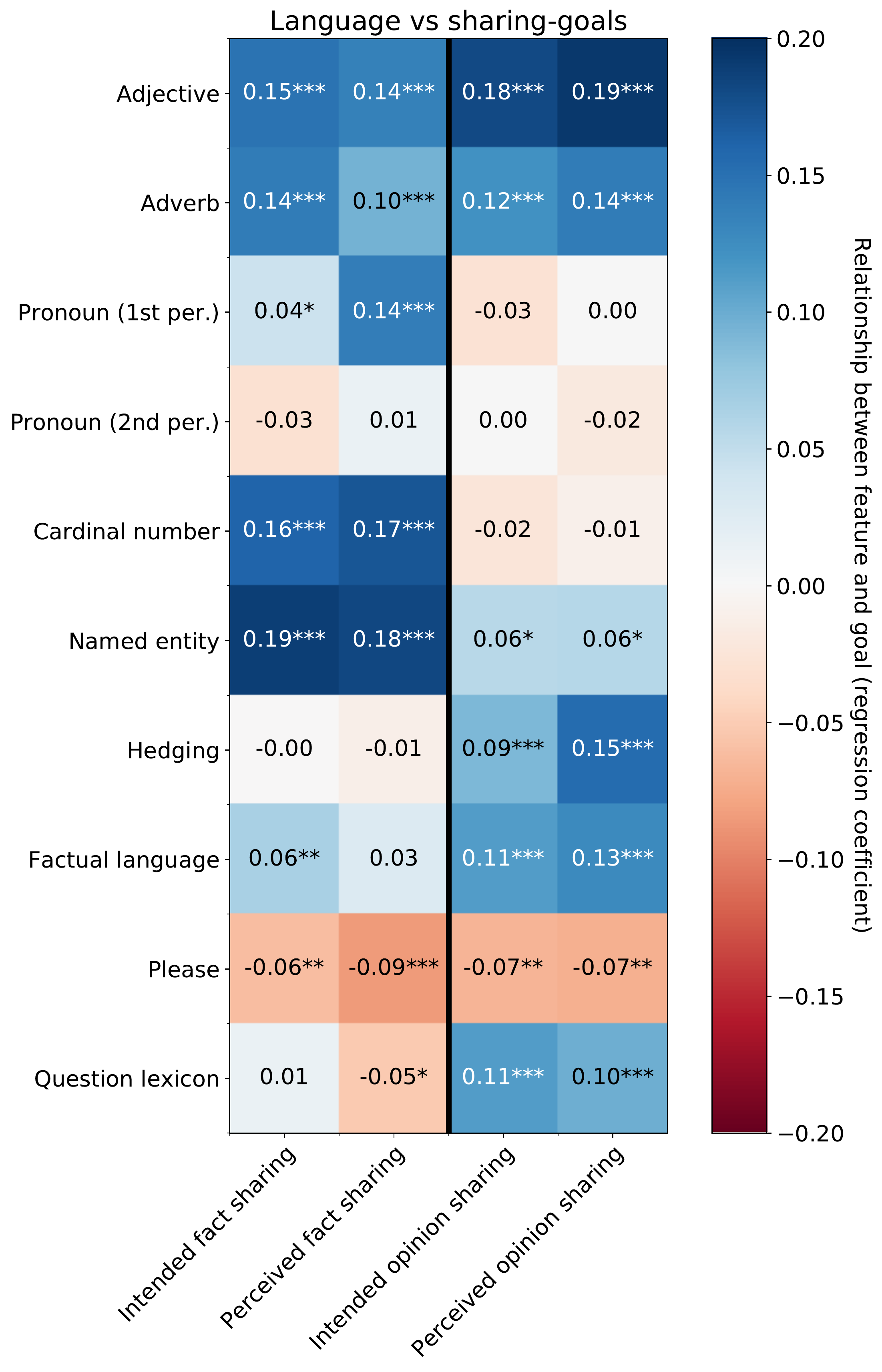}
        \caption{}
        \label{fig:sharing_language}
    \end{subfigure}
    \hfill %
    \begin{subfigure}[b]{0.48\textwidth}
        \includegraphics[width=\textwidth]{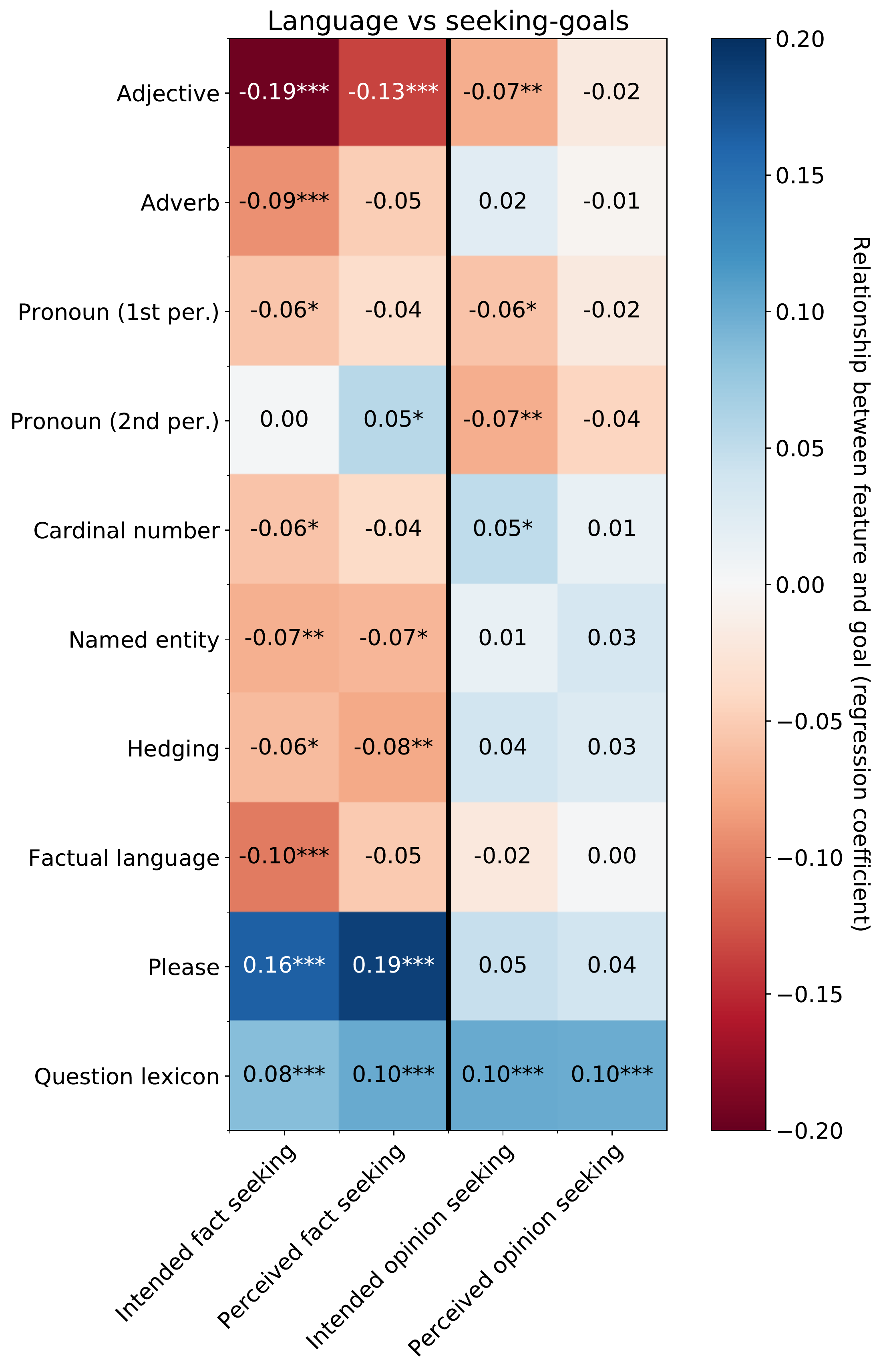}
        \caption{}
        \label{fig:seeking_language}
    \end{subfigure}
    \caption{Regression analysis comparing the rates of linguistic features in intention and perception of sharing-\intenttype{s} (a) and in intention and perception of seeking-\intenttype{s} (b). Observed correlations are generally consistent with results from literature on subjectivity detection. While most features correlate similarly between intents and perceptions, there are some that differ, including factual language and use of first or second person pronouns.}
    \label{fig:language}
\end{figure*}
\xhdr{Difference in certainty}
In several of the distributions in Figure \ref{fig:intent_perception_dists}, intention
survey
responses are often more likely to take on one of the extreme
rating options (``not at all'' or ``definitely'') compared to perception responses.
Conversely, perception responses are often more likely to display some degree of hedging by giving
ratings
of ``mostly not'', ``somewhat'', or ``mostly''.
We refer to this as a difference in certainty, and formalize it as follows: for each \intenttype, we compute the proportion of uncertain
ratings (``mostly not'', ``somewhat'', or ``mostly'')
among intention and perception responses, and compare the proportions via chi-squared test.
We take the chi-squared test statistic as the
\emph{relative uncertainty score}
for that \intenttype. A higher
 score means that the perception responses are more inclined towards uncertain
ratings
relative to the intention responses.
Ranking the \intenttype{s} by uncertainty score, we find that opinion sharing ranks lowest (uncertainty score = 7.8), which
reflects the relatively strong lean towards 5 
(``definitely'')
among reported perceptions of that \intenttype (Figure \ref{fig:opinion_giving_dist}).
Fact seeking and opinion seeking are nearly tied for the highest relative uncertainty scores (26.9 and 27.8, Figures~\ref{fig:info_seeking_dist} and~\ref{fig:opinion_seeking_dist}, respectively).
Such seeking-goals may be harder to perceive with high confidence because they are sometimes implicit \cite{kehler_evaluating_2017}.

\subsection{Linguistic cues}
\label{sec:language}
Past work found that linguistic features such as part of speech, named entities, and hedging can distinguish between subjective~\cite{wiebe_development_1999,yu_towards_2003} and objective \cite{lex_objectivity_2010,regmi_what_2015} statements (corresponding to sharing opinions and facts), and that lexicon-based features can distinguish information seeking questions (which roughly correspond to fact seeking) from other types of questions such as social coordination \cite{liu_taxonomy_2015,guy_identifying_2018,harper_facts_2009}.
But because these results relied exclusively on third-party labels, they only reflect perceptions.
Here, we explore whether these linguistic features are also indicative of intentions.

\xhdr{Selecting linguistic features}
We began with a basic set of linguistic features \cite{wiebe_development_1999}: the usage of pronouns, adjectives, cardinal numbers, modals, and adverbs.
We then refined the pronoun feature by distinguishing the use of first-person and second-person pronouns \cite{lex_objectivity_2010,harper_facts_2009}.
Mentions of named entities are characteristic of objective statements while hedging language (e.g., ``I believe...'') tends to signal subjectivity \cite{regmi_what_2015}, so we incorporated these as additional features, alongside the explicit use of factual language (e.g., ``In fact...'') which can be regarded as the opposite of hedging \cite{brown_politeness_1987}.
Finally, we added features associated with information seeking questions: the use of please \cite{guy_identifying_2018} and a question lexicon based on prior work \cite{liu_taxonomy_2015}.
For simplicity, all features were treated as binary (a comment either exhibits at least one instance of the linguistic feature or it does not).
All features were extracted from the opening comment of the conversation as that was the comment the surveys asked about.

\xhdr{Comparing linguistic features} To compare how linguistic features are tied to intentions versus perceptions, 
for each pair of linguistic feature and \intenttype (binarized, as described in Section \ref{sec:data}),
we separately regressed the intended goal on the feature, as well as the perceived goal on the feature, controlling for age, gender, Page size, and Page category.
Regression coefficients are shown in Figure \ref{fig:language}, where all variables were standardized for ease of comparison.

Several linguistic features correlate 
similarly with intentions and perceptions.
For example,
a one-standard-deviation increase in hedging corresponds to an increase in opinion sharing intent by 0.09 standard deviations and to an increase in opinion sharing intent by 0.15 standard deviations ($p < 0.001$).
Similarly, adjectives signal both intended and perceived fact sharing
(regression coefficients 0.15 and 0.14, respectively, $p < 0.001$).

Furthermore, the perception correlations generally corroborate
prior results on 
subjectivity and information seeking.
Consistent with 
findings in subjectivity detection,
mentions of named entities are more correlated with fact sharing
(0.18, $p < 0.001$)
than with opinion sharing
(0.06, $p < 0.05$),
and use of cardinal numbers (intuitively, a heuristic capturing mentions of specific values) is correlated with fact sharing
(0.17, $p < 0.001$)
and not opinion sharing.
Conversely, hedging is 
correlated with opinion sharing
(0.15, $p < 0.001$)
but not fact sharing.
Consistent with prior work on information seeking questions, the use of please is associated with fact seeking (0.19, $p < 0.001$) and not opinion seeking;
a similar but weaker effect holds for second person pronouns (0.05, $p < 0.05$).

Although many of the observed correlations are largely similar between intentions and perceptions, there are 
also some 
notable differences.
For instance, the use of factual language is significantly correlated with intended fact sharing
(0.06, $p < 0.01$)
but not with perceived fact sharing.
This may relate to the previously observed bias towards perceiving statements as opinions: even if the initiator tries to ``double down'' on the intended factuality of their comment through the explicit use of factual language, this might not have any effect on the replier, who is inclined towards perceiving opinions.
Other examples include
question words being negatively correlated with perceived fact sharing (-0.05, $p < 0.05$) but uncorrelated with intended fact sharing,
and second person pronouns being correlated with perceived
(0.05, $p < 0.05$)
but not with intended fact seeking.
Future work could examine in greater detail why these differences occur, and also consider more sophisticated language features.

\begin{figure*}
    \centering
    \begin{subfigure}[b]{0.48\textwidth}
        \includegraphics[width=\textwidth]{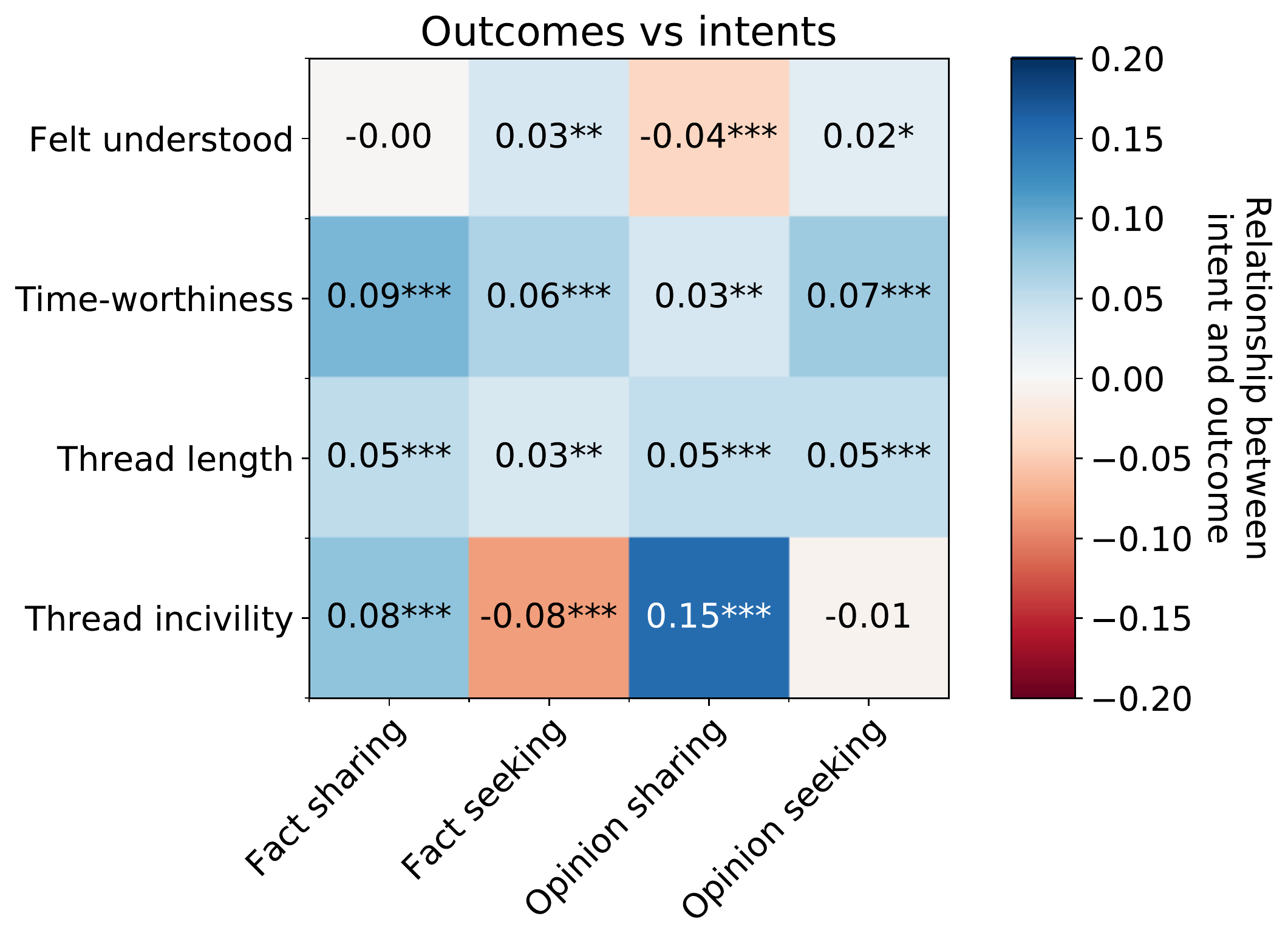}
        \caption{}
        \label{fig:intent_regressions}
    \end{subfigure}
    \hfill %
    \begin{subfigure}[b]{0.48\textwidth}
        \includegraphics[width=\textwidth]{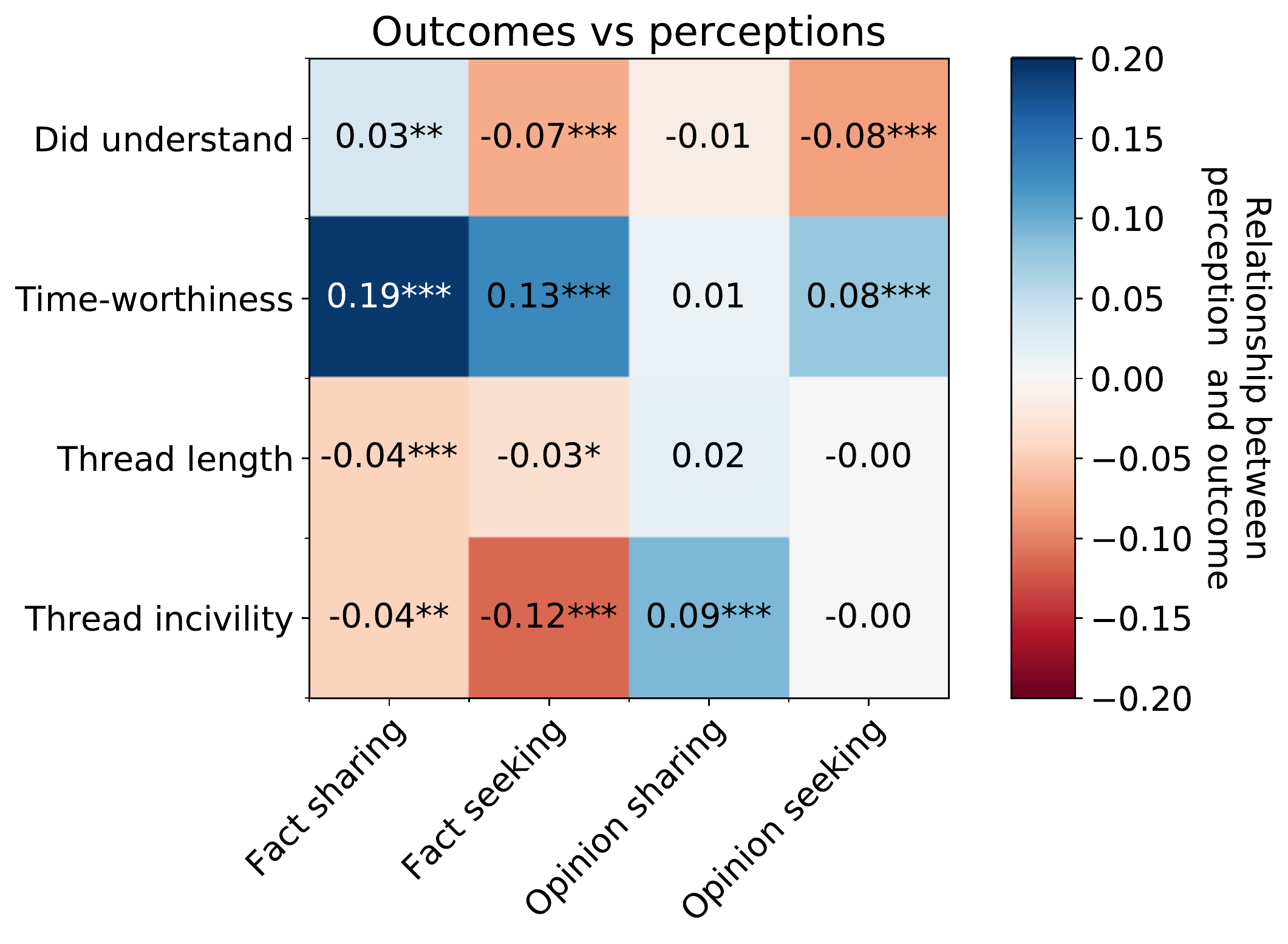}
        \caption{}
        \label{fig:perception_regressions}
    \end{subfigure}
    \caption{Conversational outcomes can relate differently to intentions (a) and perceptions (b). For example, while an intention to share a fact is positively correlated with greater future incivility in the conversation (regression coefficient 0.08), a perception that a comment is sharing a fact is instead negatively correlated (-0.04).}
    \label{fig:regressions}
\end{figure*}

\xhdr{Predicting conversational intentions and perceptions}
Linguistic features, such as the ones described in the preceding analysis, have previously been successfully used in models for predicting subjectivity from text \cite{wiebe_learning_2004,lex_objectivity_2010,yu_towards_2003}.
However, these models' reliance on third-party labels means that they must be understood as predicting \emph{perceived} subjectivity.
We now leverage our unique access to ground truth intentions to address the following question: Can intentions also be predicted from text, and if so, how different are intention prediction models from perception prediction models?

\begin{table}[t]
    \begin{tabular}{rcc|cc|cc}
        \toprule
        ~ & \multicolumn{2}{c|}{\textit{Perception}} & \multicolumn{2}{c|}{\textit{Intention}} & \multicolumn{2}{c}{\textit{Label}} \\
        ~ & \multicolumn{2}{c|}{\textit{prediction}} & \multicolumn{2}{c|}{\textit{prediction}} & \multicolumn{2}{c}{\textit{swap}} \\
        \textbf{Goal} & \textbf{IC} & \textbf{+R} & \textbf{IC} & \textbf{+R} & \textbf{IC} & \textbf{+R} \\
        \midrule
        Fact sharing & .64 & .64 & .68 & .68 & .65 & .64 \\
        Fact seeking & .81 & .83 & .82 & .83 & .80 & .80 \\
        Opinion sharing & .72 & .74 & .75 & .78 & .75 & .77 \\
        Opinion seeking & .64 & .65 & .72 & .71 & .68 & .68 \\
        \midrule
    \end{tabular}
    \caption{BERT-based classifiers using either the text of the initiator's comment only (IC) or the text of both the initiator's comment and the reply (+R) achieve reasonable performance in predicting both perceptions (left) and intentions (middle).
    Furthermore, using perception labels to predict intentions (``label swap'') results in performance drops compared to using intention labels (compare middle and rightmost columns).
    All results are reported as area under the ROC curve 
    (AUC) to account for class imbalance.}
    \label{tab:intent_perception_preds}
\end{table}

To evaluate the feasibility of intention prediction, we fine-tuned a BERT-based classifier \cite{devlin_bert_2019} on the task of predicting the initiator's intention based on the text of their initial comment. Since prior work has found that incorporating context can improve predictive performance in conversational settings \cite{gao_detecting_2017,ghosh_role_2017}, for completeness we additionally considered a version of this model that also looks at the text of the reply. Both models were trained on about 5,000 samples from the intention survey data, using binarized intention responses as the labels. They were then evaluated on 1,000 held out test samples from the same survey. Finally, as a point of comparison, we also trained models for the more traditional task of perception prediction by using the same setup on the perception survey data.

Table \ref{tab:intent_perception_preds} (leftmost two columns) compares the performance of the intention and perception classifiers, measured in terms of area under the ROC curve (AUC) to account for class imbalance. We find that both intentions and perceptions can be predicted with similar performance, thus establishing the feasibility of the intention prediction task.

Beyond demonstrating feasibility, we also want to understand if predicting intention differs from predicting perception.
In other words, are 
intention labels provided by the authors themselves interchangeable with ``third-party'' labels, not unlike those used in prior work on subjectivity detection?
To 
test
this, we applied the 
model trained on perception labels to the intention-labeled test set.
We find that using perception labels to predict intentions (Table~\ref{tab:intent_perception_preds}, rightmost two columns) results in reduced performance for all four \intenttype{s} compared to using intention labels, with a 3.4\% average decrease in AUC.
One possible explanation for the difference is that models trained on perception labels may be learning the (distributional and linguistic) perception biases described earlier.
If so, the use of such models should account for this limitation, especially in production settings.

\subsection{Relationship to conversational outcomes}
\label{sec:outcomes}
Intentions and perceptions can also differ in their relationship to the outcomes and trajectories of the conversations in which they occur.
We considered several conversational properties, three of which rely on the survey responses:
whether the initiator or replier felt the discussion was worth their time,
whether the initiator felt understood,
or whether the replier felt they understood the initiator.
We also considered two conversational trajectories
proposed in prior work:
the eventual length of the thread \cite{backstrom_characterizing_2013,aragon_generative_2017,kumar_dynamics_2010}
and whether the discussion eventually turns uncivil \cite{liu_forecasting_2018}.
To formalize the latter outcome at scale, we defined \emph{thread incivility} as the maximum incivility score of all comments in the conversation following the initial comment-reply pair, where incivility score is computed by a production DeepText DocNN classifier \cite{collobert_natural_2011,kim_convolutional_2014,zhang_text_2015} trained on manually-labeled content that violates Facebook's Community Standards on Hate Speech.\footnote{\url{https://www.facebook.com/communitystandards/hate_speech}}
We verified the reliability of these 
scores by manually annotating a random sample of 200 comments using prior guidelines \cite{zhang_conversations_2018}.
Substantial agreement between the manual and automated labels (Cohen's $\kappa$ = 0.73) suggests that this automated score is a reasonable 
measure of incivility.

\xhdr{Intentions and outcomes}
As before, we compared each pair of goal and outcome using a controlled regression analysis, regressing outcome on goal.
Results are shown in Figure \ref{fig:intent_regressions}.

\noindent\emph{Opinion sharing.}
An intention to share an opinion is correlated with higher likelihood of future incivility (regression coefficient 0.15, $p < 0.001$).
This corroborates past work suggesting that opinion sharing is correlated with flaming \cite{moor_flaming_2010}.
The intention to share opinions is also correlated with stronger feelings of being misunderstood (-0.04, $p < 0.001$)
and with longer threads (0.05, $p < 0.001$); the latter may be the result of opinion sharing 
triggering extended arguments or debates.
However, initiators also tend to rate conversations started with opinion sharing intent as being worth their time (0.03, $p < 0.01$).
One possible explanation is that initiators perceive the act of sharing their opinion as inherently valuable,
regardless of
downstream interactional outcomes.
Together, these observations suggest a potential reason for why
incivility continues
to be prevalent on many online discussion platforms: people feel that sharing their opinion is worth their time despite the increased likelihood that doing so will lead to undesirable outcomes.
These observations motivate further work on better understanding why conversation participants rate interactions as worth their time
and on the design of platforms that can offer an outlet for expressing personal opinions while also encouraging healthy conversations around those opinions.

\noindent\emph{Seeking versus sharing facts.}
Among all the \intenttype{s}, fact seeking appears to be the most unambiguously positive, being significantly correlated with
lower thread incivility (-0.06, $p < 0.001$), feeling understood (0.03, $p < 0.01$), and considering the conversation to be worth the time (0.06, $p < 0.001$).
This could suggest that in
many
 of these cases the initiator ends up getting the information they sought, leading them to view the interaction positively.
On the other hand, fact sharing is
slightly associated with negative outcomes: like opinion sharing, it is positively correlated with thread incivility (0.06, $p < 0.001$),
although, unlike opinion sharing, it is not related to
  feeling understood.

\xhdr{Perceptions and outcomes}
Several of the correlations we observed for intentions also hold for perceptions (Figure~\ref{fig:perception_regressions}).
Notably, perceived opinion sharing remains correlated with
higher thread incivility (0.09, $p < 0.001$)
while perceived fact seeking remains correlated with
lower thread incivility (-0.12, $p < 0.001$).

However, perceptions and intentions relate differently to outcomes in some key ways.
In particular, for fact sharing, the direction of the correlation flips for both thread length and thread incivility.
When an initial comment is \emph{intended} to share a fact, the resulting conversation is more likely to turn uncivil (0.08, $p < 0.001$) and tends to run longer (0.05, $p < 0.001$).
When an initial comment is \emph{perceived} to be sharing a fact, the resulting conversation is less likely to turn uncivil (-0.05, $p < 0.01$) and tends to run shorter (-0.04, $p < 0.001$).
Since this contrast might provide new insights into why some online public discussions turn uncivil, we examine it in more detail in the following section.

\section{Misperception of Fact Sharing: \\ A Case Study}
\label{sec:eval}

So far, we revealed systematic differences between intentions and perceptions at an aggregate level.
In this section, we investigate the effect of \emph{misperception} at the discussion level, or what happens in a conversation in which a replier perceives an initiator's comment differently from how it was intended.
Particularly, we examine the relationship between misperception and undesirable conversational outcomes such as future incivility in a discussion.

As discussed above (Section~\ref{sec:outcomes}) intended fact sharing in the initial comment correlates with greater incivility later in the discussion, but perceived fact-sharing instead correlates with less incivility later on---could misperception explain this contrast?

As facts are often misperceived as opinions \cite{rabinowitz_distinguishing_2013,mitchell_distinguishing_2018}, we suspect perceptions of opinion-sharing may play a role. This, combined with the additional observation that perceived opinion sharing correlates with greater incivility, leads us to hypothesize:
\emph{the observed positive correlation between intended fact sharing and thread incivility can be attributed to comments that, while intended to share a fact, get misperceived as
sharing an opinion}.

Testing this hypothesis requires labels for both intention and perception on the same conversations, but obtaining such paired ground truth at scale is infeasible (see Section~\ref{sec:data}).
To circumvent this limitation, we explored two alternative approaches for obtaining paired labels.
Our first approach supplements ground truth intention labels with automatically inferred perception labels, exploiting the relatively good performance of our perception classifiers (Section \ref{sec:language}).
Still, these classifiers may not (mis)perceive comments the same way that humans do, so any findings may simply reflect classifier error rather than human misperception.
As such, our second approach combines the ground truth intention labels with third-party human-annotated perception labels, albeit only on a random subset of data due to platform limitations.
Each approach has its drawbacks, but both lead to the same qualitative conclusion that supports our hypothesis.

\subsection{Automatically inferred perceptions}

\xhdr{Labeling procedure}
We started by finding all comments whose ground truth intentions were, according to the initiator survey responses, to share a fact and \emph{not} an opinion.
Cases of mixed intention were excluded as they make misperception ambiguous: if a comment intended as sharing both fact and opinion is perceived as only sharing opinion, is it correctly perceived (as the perceiver correctly inferred the opinion sharing intent) or misperceived (as the perceiver failed to infer the fact sharing intent)?

We then ran the perception classifier (Section \ref{sec:language}) on these comments.%
\footnote{We use the version that uses text from both the initial comment and the reply as it performs best.}
For each comment, the classifier returns a confidence score between 0 and 1 for each \intenttype, representing the estimated likelihood that the initiator's comment in the conversation was perceived as having that \intenttype.
The result is a dataset of comments where each comment has a fixed ground truth intention (fact sharing and not opinion sharing) and two classifier-generated perception scores,
one for perceived opinion sharing and one for perceived fact sharing (the latter is included as a control).
The classifier can be thought of as an
imperfect proxy for a
human perceiver.

\xhdr{Method}
To verify our hypothesis, we regressed future incivility on perceived opinion sharing while accounting for several possible confounds.
First, we controlled for perceived fact sharing (and included an interaction effect with perceived opinion sharing) to test the alternate hypothesis that perceived fact sharing alone fully explains differences in incivility.
We also controlled for the incivility of the initial comment and that of the reply, as prior work found that incivility in the opening exchange of a conversation is a relatively strong indicator of future incivility \cite{zhang_conversations_2018}.
Finally, we also included the gender and age of both the initiator and the replier, Page size, and Page category.
All continuous variables were standardized.

If our hypothesis holds, perceived opinion sharing would be positively associated with future incivility.

\begin{table}
  \renewcommand*{\arraystretch}{1.15}
  \centering
  \begin{tabular*}{\columnwidth}{lrrl}\\
    \multicolumn{4}{c}{{\textbf{Future incivility}}} \\
    {\textit{Predictor}} &
      {\textit{\textbeta}} &
      {\textit{SE}} & \\
    \midrule
    (Intercept) &
      0.37 & 0.63 & \\
    Perceived fact sharing (predicted) &
      -0.12 & 0.05 & * \\
    \textbf{Perceived opinion sharing (predicted)} &
      \textbf{0.30} & \textbf{0.05} & *** \\
    Perceived fact sharing $\times$ opinion sharing &
      0.04 & 0.05 & \\
    Initial comment incivility &
      0.07 & 0.04 & $\hat{}$ \\
    Reply incivility &
      0.19 & 0.05 & *** \\
    Initiator is female &
      -0.31 & 0.09 & *** \\
    Replier is female &
      -0.01 & 0.08 & \\
    Initiator age &
      -0.03 & 0.04 & \\
    Replier age &
      0.10 & 0.04 & * \\
    Page size (logged) &
      0.03 & 0.04 & \\
    Page category (not shown for space) &
      & & \\
    \bottomrule
  \end{tabular*}
  \caption{A regression analysis reveals the relationship between misperception and incivility in discussions where the initial comment was intended to share a fact ($R^2 = 0.23$). (*** \textit{p} < 0.001, ** \textit{p} < 0.01, * \textit{p} < 0.05, $\hat{}$ \textit{p} < 0.10).}~\label{tab:regressions}
\end{table}

\xhdr{Results}
Consistent with our hypothesis, we find a significant positive effect of perceived opinion sharing (\textit\textbeta = 0.30, $p < 0.001$): a one-standard-deviation increase in perceived opinion sharing results in a 0.30 standard-deviation increase in future incivility (Table \ref{tab:regressions}).
This effect dominates the effect of other variables in the regression except that of the initiator's gender, which is about equal in magnitude.\footnote{Analysis of the relationship between gender and incivility lies outside the scope of the present work; see \cite{craker_dark_2016} for some additional discussion.}
Consistent with our previous findings, we also find a significant but weaker negative effect of perceived fact sharing (\textit\textbeta = -0.12, $p < 0.05$), indicating that correctly perceiving fact sharing
translates into more civil conversations.

\begin{figure}
    \centering
    \includegraphics[width=0.9\linewidth]{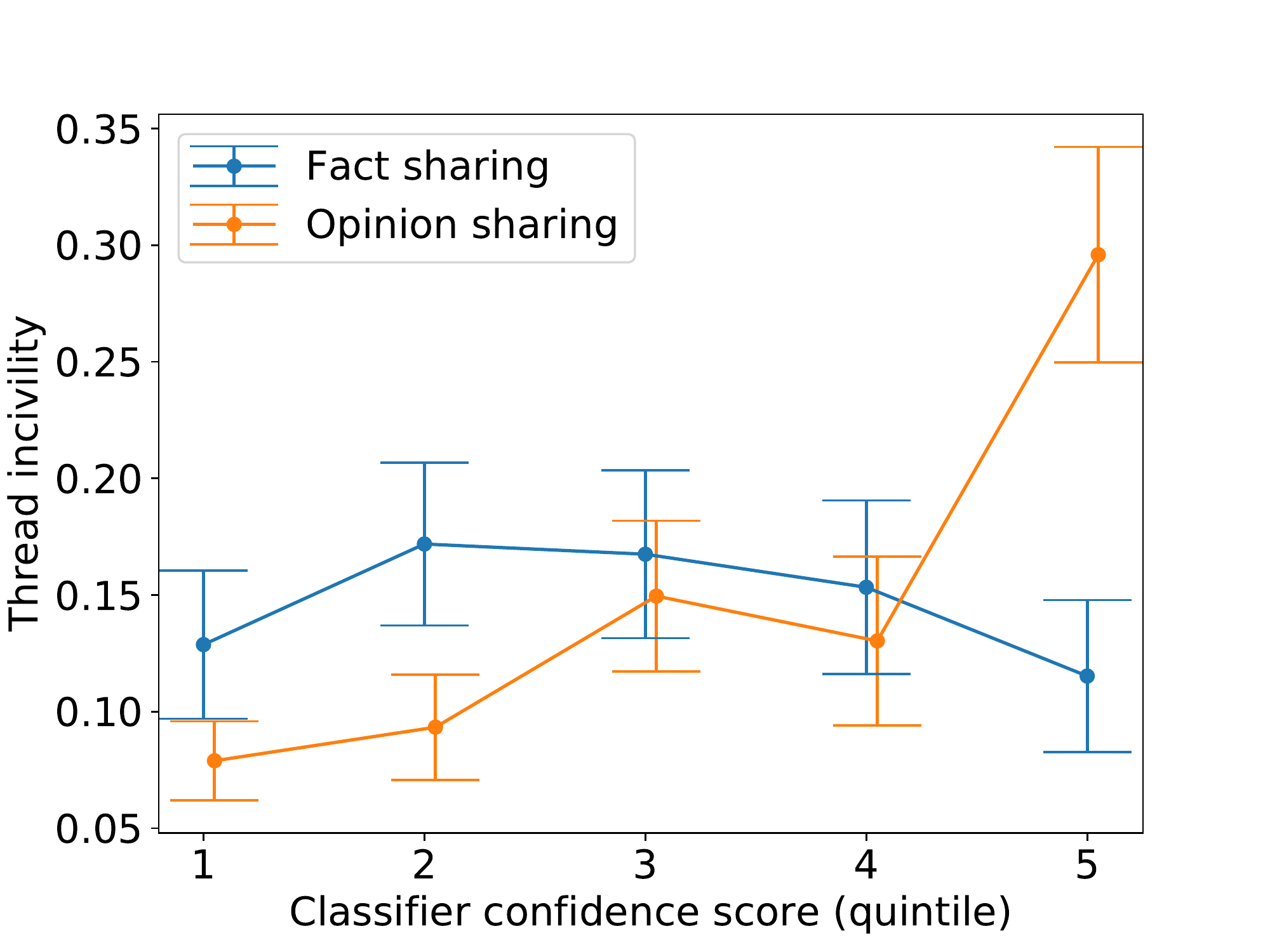}
    \caption{Visualization of the relationship between thread incivility and perception scores for fact sharing (blue) and opinion sharing (orange), conditioned on the ground-truth intent being fact sharing. Each point indicates mean thread incivility among all comments in the specified quintile of classifier scores; error bars indicate 95\% CIs.}
    \label{fig:ig-perceived-og}
\end{figure}

These relationships can also be visualized (in an uncontrolled setting) by binning conversations by quintiles of the perception classifier scores for fact sharing and opinion sharing, then plotting mean incivility per bin (Figure \ref{fig:ig-perceived-og}). Again, we observe a positive relationship between perceived opinion sharing and thread incivility
and a weak negative relationship between perceived fact sharing and thread incivility.

\subsection{Human annotated perceptions}
Automatically inferred perceptions have the advantage of scalability but the findings made using those labels may only apply to cases where a classifier misperceives fact sharing as opinion sharing, and may not necessarily generalize to cases where a person does the same.
As such, we also obtained third-party human annotations of perception to validate the previous results.

\xhdr{Labeling procedure}
We sent a random sample of conversations from the initiator survey responses to
expert annotators (though platform limitations restricted the scale of annotation).
To avoid potentially biasing the annotators, comments were sampled such that half were intended as fact sharing and half were intended as opinion sharing.
As before, the two intentions were held to be mutually exclusive.
For each conversation, annotators were shown the initial comment and asked to rate whether they thought it was sharing a fact and, separately, whether they thought it was expressing an opinion (such that they had the option of marking both or neither).
The terms ``opinion'' and ``fact'' were left purposely vague and annotators were encouraged to exercise personal judgment, to more accurately simulate how perceptions get formed upon seeing a comment in the wild.
In total, we received annotations for 330 comments from 6 annotators.

\xhdr{Method}
Due to the smaller data size and coarser-grained labels, we ran a simplified analysis, comparing the future incivility of discussions in which the initial comment was labeled as expressing an opinion to discussions in which the initial comment was labeled as sharing a fact.
If our hypothesis holds, we expect future incivility to be higher in the former case.

\xhdr{Results}
Supporting this hypothesis, we find that the median thread incivility among cases labeled as expressing opinions is 0.09, compared to 0.07 among cases labeled as sharing facts; this difference is significant via Mann-Whitney test ($U = 1655.5$, $p < 0.05$).
This finding also provides additional evidence that the differences we have observed in this section are actually reflective of differences in perception, as opposed to merely reflecting biases of the classifier.

Together, the results from both perception label sources (automated and human) provide support for our hypothesis, suggesting that among conversations intended as fact sharing, the correlation with thread incivility arises largely from cases that were (mis)perceived as opinion sharing.
Future work could build upon this result by investigating what factors (linguistic or otherwise) lead to this kind of misperception, and whether similar effects occur for other combinations of \intenttype{s} and outcomes.
This line of research could lead to a better understanding of the mechanisms through which incivility arises in well-intentioned online discussions.

\section{Further Related Work}
\label{sec:related}
\xhdr{Intentions in other domains}
While our work focused on conversational intentions, intentionality is also a dimension of other online settings.
For instance, prior work studied intent to communicate (offline) commitments in emails \cite{carvalho_modeling_2011,wang_context-aware_2019,lin_actionable_2018}, search query intent for search result customization \cite{hanjalic_intent_2012,ashkan_classifying_2009,brenes_survey_2009,hashemi_query_2016}, and purchasing intent on online shopping platforms \cite{brown_buying_2003,ling_effects_2010,heijden_predicting_2001}.
These settings differ from our setting of online discussions, but there is some overlap in \intenttype{s}: fact seeking intent also applies to email \cite{wang_context-aware_2019} and search \cite{broder_taxonomy_2002}.

Some forms of intention can be measured without the need for surveys.
For instance, some work has studied intended humor \cite{reyes_humor_2012} and sarcasm \cite{bamman_contextualized_2015,gonzalez-ibanez_identifying_2011} by treating user-supplied hashtags as natural labels.
In the context of news articles, document tags were treated as natural labels of the intended purpose of the tagged article \cite{yu_towards_2003}.
Linguistic features such as part-of-speech
ended up being predictive of both these natural intent labels \cite{bamman_contextualized_2015} and our survey-based labels.

\xhdr{Factors influencing perception}
Like intention, perception is also relevant in many online settings, and prior work has explored factors influencing how online comments, documents, and actions get perceived.
In particular, prior work on the perceived objectivity of online news \cite{sundar_effect_1998} and the perceived trustworthiness of product reviews \cite{filieri_what_2016} or dating profiles \cite{jin_match_2015} closely relate to our work on perception of facts and opinions, while studies of how perceived fairness of community moderation affects future incivility or community loyalty \cite{jhaver_did_2019,chang_trajectories_2019} echo our results relating perceptions to conversational outcomes.
We add to the existing literature on perception by using our survey-based methodology to relate perceptions to intentions.
Most similar is previous work that surveyed 95 conversation initiators and 41 repliers to measure intentions and perceptions in relation to incivility \cite{moor_flaming_2010}.
The present work, in contrast, involved a much larger-scale survey and considered other outcomes such as thread length.

\xhdr{Subjectivity detection}
Distinguishing between opinions and facts is closely related to the task of subjectivity detection, for which a number of language-based models have been proposed \cite{wiebe_learning_2004,wiebe_development_1999,lin_sentence_2011,murray_subjectivity_2011}; see \cite{liu_sentiment_2010} for a more complete survey.
However, the two tasks are not identical as subjectivity encompasses more than opinions: \cite{wiebe_development_1999} defines subjective language as expressing private state, which includes not only opinions but also emotions and speculation \cite{quirk_comprehensive_1985}.
Some work on subjectivity focused explicitly on opinions in the context of news media and wiki articles \cite{regmi_what_2015,wiebe_recognizing_2003}%
, but largely relied on third-party annotations \cite{wiebe_annotating_2005} and hence mainly captured perceptions of opinions.
We add to this work by
examining opinions and facts in a conversational context (which in turn introduces a distinction between sharing and seeking), considering intentions to share opinions (or facts) in addition to perceptions, and comparing the use of linguistic cues borrowed from the subjectivity detection literature in predicting intentions and perceptions.

Subjectivity detection has also been shown to be helpful in downstream tasks such as information extraction \cite{riloff_exploiting_2005}, sentiment analysis \cite{yessenalina_multi-level_2010}, and document quality measurement \cite{lex_measuring_2012}.
Our work similarly shows that both intention and perception of opinions and facts can be indicative of conversational outcomes such as future incivility.

\xhdr{Forecasting conversational outcomes}
This work has shown that intentions and perceptions relate to future conversational outcomes.
Prior work has studied other signals of outcomes, such as pragmatic cues \cite{zhang_conversations_2018}, similarity between comments \cite{althoff_large-scale_2016}%
, and conversation structure \cite{garimella_quantifying_2017}.
These have been used to forecast outcomes such as success in negotiation \cite{curhan_thin_2007,cadilhac_grounding_2013} and eventual disagreement \cite{hessel_somethings_2019}, as well as two of the outcomes examined in our work: thread length \cite{backstrom_characterizing_2013} and incivility
\cite{zhang_conversations_2018,liu_forecasting_2018}.
In particular, \cite{zhang_conversations_2018} indirectly explores the connection between intentions and future incivility by using an unsupervised  method~\cite{zhang_asking_2017} to estimate the intended role of a comment%
; we build on this by obtaining ground truth intents via survey and additionally relating them to perceptions.

\section{Conclusion}
\label{sec:discussion}

In this work, we presented a large-scale study of how intentions and perceptions can diverge in online public discussions.
Using a survey of over 16,000 people, we obtained unprecedented access to ground truth labels for the intentions underlying comments on online public discussions, as well as how such comments were perceived.
Using this data, we revealed both distributional and linguistic differences between intentions and perceptions, showed that such differences are reflected in the performance of automated classifiers, and explored how misperceptions can be tied to the future trajectory of a discussion.
In particular, when a comment intended to share a fact is misperceived as sharing an opinion, the subsequent conversation is more likely to turn uncivil than when that intention is correctly perceived.

These results point towards several design opportunities for promoting healthier interactions on online discussion platforms.
For instance, classifiers that predict intentions and perceptions could signal to users when a comment they are writing may be misperceived by others and suggest concrete strategies for reducing this risk.
Nonetheless, user studies would be needed to guide the design of such interventions to reduce the likelihood of unintended negative consequences.
Our results further suggest that reducing misperception may improve civility in online discussions, but additional work is needed to better understand this connection and to what extent, if at all, misperception is \emph{predictive} of incivility.
We also note that these findings specifically apply to public discussions, and the effect of misperception in other kinds of settings (e.g., private discussions or article-style monologic text) remains a related but separate question.

Limitations of our survey methodology provide opportunities for future work.
Low survey response rates
prevented the collection of paired survey responses from an initiator and replier on the same conversation, limiting a direct study of misperception.
The classifier predictions we used as a substitute for perception labels were generally reliable, and our results were verified via third-party annotation, but it would nonetheless be valuable to replicate these results on paired responses.
The retrospective nature of the surveys also adds ambiguity with respect to interpretations of reported perceptions: were responders reporting how they perceived the comment at the time of the conversation, or were they reporting how they perceived it in hindsight at the time of the survey?
Though the results varied little with the amount of time between the time a person commented and the time they took the survey, surveying people at the time of conversation may constitute interesting future work.
Finally, while we have accounted for demographic and Page features as potential confounds, other confounds may exist.
Our analysis may also be applied to other conversational \intenttype{s} beyond facts and opinions.
For instance, community moderators may want to deal differently with a person who intentionally trolled others in a conversation than one who unintentionally did so.
Intentions and perceptions may also relate to community-level rather than conversation-level outcomes.
For instance, does the way in which a community member tends to perceive others in the community relate to that member's long-term loyalty?
Such effects might also end up being specific to certain \emph{kinds} of communities; while our present work looks only at Facebook Pages, future work could apply our methodology to other platforms with different modes and norms of interaction, like Twitter and Reddit.
Finally, while our present work has looked at intentions and perceptions at the start of a conversation, \intenttype{s} may change as the conversation progresses.
These all constitute promising paths for future exploration building upon the methods and findings presented in this work.\footnote{All analysis code is available on Github at \url{https://github.com/facebookresearch/intentions-perceptions}.}

\begin{acks}
This research was supported in part by an NSF CAREER award IIS-1750615 and by an NSF Grant IIS-1910147.
We would like to thank Lada Adamic, Israel Nir, Alex Dow, Ashish Gupta, Karen Jusko, Alex Leavitt, Moira Burke, Caleb Chiam, Liye Fu, Justine Zhang, as well as our reviewers for their valuable feedback (which we hopefully perceived correctly).

\end{acks}

\bibliographystyle{ACM-Reference-Format}
\bibliography{intents-www-autoupdate-jpc,intents-www-autocomplete-C}


\begin{thebibliography}{71}


\ifx \showCODEN    \undefined \def \showCODEN     #1{\unskip}     \fi
\ifx \showDOI      \undefined \def \showDOI       #1{#1}\fi
\ifx \showISBNx    \undefined \def \showISBNx     #1{\unskip}     \fi
\ifx \showISBNxiii \undefined \def \showISBNxiii  #1{\unskip}     \fi
\ifx \showISSN     \undefined \def \showISSN      #1{\unskip}     \fi
\ifx \showLCCN     \undefined \def \showLCCN      #1{\unskip}     \fi
\ifx \shownote     \undefined \def \shownote      #1{#1}          \fi
\ifx \showarticletitle \undefined \def \showarticletitle #1{#1}   \fi
\ifx \showURL      \undefined \def \showURL       {\relax}        \fi
\providecommand\bibfield[2]{#2}
\providecommand\bibinfo[2]{#2}
\providecommand\natexlab[1]{#1}
\providecommand\showeprint[2][]{arXiv:#2}

\bibitem[\protect\citeauthoryear{Althoff, Clark, and Leskovec}{Althoff
  et~al\mbox{.}}{2016}]%
        {althoff_large-scale_2016}
\bibfield{author}{\bibinfo{person}{Tim Althoff}, \bibinfo{person}{Kevin Clark},
  {and} \bibinfo{person}{Jure Leskovec}.} \bibinfo{year}{2016}\natexlab{}.
\newblock \showarticletitle{Large-Scale {{Analysis}} of {{Counseling
  Conversations}}: {{An Application}} of {{Natural Language Processing}} to
  {{Mental Health}}}.
\newblock \bibinfo{journal}{\emph{Transactions of the Association for
  Computational Linguistics}} (\bibinfo{date}{Dec.} \bibinfo{year}{2016}).
\newblock


\bibitem[\protect\citeauthoryear{Arag{\'o}n, G{\'o}mez, Garc{\'i}a, and
  Kaltenbrunner}{Arag{\'o}n et~al\mbox{.}}{2017}]%
        {aragon_generative_2017}
\bibfield{author}{\bibinfo{person}{Pablo Arag{\'o}n}, \bibinfo{person}{Vicen{\c
  c} G{\'o}mez}, \bibinfo{person}{David Garc{\'i}a}, {and}
  \bibinfo{person}{Andreas Kaltenbrunner}.} \bibinfo{year}{2017}\natexlab{}.
\newblock \showarticletitle{Generative Models of Online Discussion Threads:
  State of the Art and Research Challenges}.
\newblock \bibinfo{journal}{\emph{Journal of Internet Services and
  Applications}} \bibinfo{volume}{8}, \bibinfo{number}{15}
  (\bibinfo{date}{Dec.} \bibinfo{year}{2017}).
\newblock


\bibitem[\protect\citeauthoryear{Artzi, Pantel, and Gamon}{Artzi
  et~al\mbox{.}}{2012}]%
        {artzi_predicting_2012}
\bibfield{author}{\bibinfo{person}{Yoav Artzi}, \bibinfo{person}{Patrick
  Pantel}, {and} \bibinfo{person}{Michael Gamon}.}
  \bibinfo{year}{2012}\natexlab{}.
\newblock \showarticletitle{Predicting Responses to Microblog Posts}. In
  \bibinfo{booktitle}{\emph{Proceedings of {{NAACL}}}}.
\newblock


\bibitem[\protect\citeauthoryear{Ashkan, Clarke, Agichtein, and Guo}{Ashkan
  et~al\mbox{.}}{2009}]%
        {ashkan_classifying_2009}
\bibfield{author}{\bibinfo{person}{Azin Ashkan}, \bibinfo{person}{Charles L.~A.
  Clarke}, \bibinfo{person}{Eugene Agichtein}, {and} \bibinfo{person}{Qi Guo}.}
  \bibinfo{year}{2009}\natexlab{}.
\newblock \showarticletitle{Classifying and {{Characterizing Query Intent}}}.
  In \bibinfo{booktitle}{\emph{Advances in {{Information Retrieval}}}},
  \bibfield{editor}{\bibinfo{person}{Mohand Boughanem},
  \bibinfo{person}{Catherine Berrut}, \bibinfo{person}{Josiane Mothe}, {and}
  \bibinfo{person}{Chantal {Soule-Dupuy}}} (Eds.).
  \bibinfo{publisher}{{Springer Berlin Heidelberg}}.
\newblock


\bibitem[\protect\citeauthoryear{Backstrom, Kleinberg, Lee, and
  {Danescu-Niculescu-Mizil}}{Backstrom et~al\mbox{.}}{2013}]%
        {backstrom_characterizing_2013}
\bibfield{author}{\bibinfo{person}{Lars Backstrom}, \bibinfo{person}{Jon
  Kleinberg}, \bibinfo{person}{Lillian Lee}, {and} \bibinfo{person}{Cristian
  {Danescu-Niculescu-Mizil}}.} \bibinfo{year}{2013}\natexlab{}.
\newblock \showarticletitle{Characterizing and {{Curating Conversation
  Threads}}: {{Expansion}}, {{Focus}}, {{Volume}}, {{Re}}-Entry}. In
  \bibinfo{booktitle}{\emph{Proceedings of {{WSDM}}}}.
\newblock


\bibitem[\protect\citeauthoryear{Bamman and Smith}{Bamman and Smith}{2015}]%
        {bamman_contextualized_2015}
\bibfield{author}{\bibinfo{person}{David Bamman} {and} \bibinfo{person}{Noah~A.
  Smith}.} \bibinfo{year}{2015}\natexlab{}.
\newblock \showarticletitle{Contextualized {{Sarcasm Detection}} on
  {{Twitter}}}. In \bibinfo{booktitle}{\emph{Proceedings of {{ICWSM}}}}.
\newblock


\bibitem[\protect\citeauthoryear{Brenes, {Gayo-Avello}, and
  {P{\'e}rez-Gonz{\'a}lez}}{Brenes et~al\mbox{.}}{2009}]%
        {brenes_survey_2009}
\bibfield{author}{\bibinfo{person}{David~J. Brenes}, \bibinfo{person}{Daniel
  {Gayo-Avello}}, {and} \bibinfo{person}{Kilian {P{\'e}rez-Gonz{\'a}lez}}.}
  \bibinfo{year}{2009}\natexlab{}.
\newblock \showarticletitle{Survey and Evaluation of Query Intent Detection
  Methods}. In \bibinfo{booktitle}{\emph{Proceedings of the 2009 Workshop on
  {{Web Search Click Data}}}}.
\newblock


\bibitem[\protect\citeauthoryear{Broder}{Broder}{2002}]%
        {broder_taxonomy_2002}
\bibfield{author}{\bibinfo{person}{Andrei Broder}.}
  \bibinfo{year}{2002}\natexlab{}.
\newblock \showarticletitle{A Taxonomy of Web Search}.
\newblock \bibinfo{journal}{\emph{ACM SIGIR Forum}} \bibinfo{volume}{36},
  \bibinfo{number}{2} (\bibinfo{date}{Jan.} \bibinfo{year}{2002}).
\newblock


\bibitem[\protect\citeauthoryear{Brown, Pope, and Voges}{Brown
  et~al\mbox{.}}{2003}]%
        {brown_buying_2003}
\bibfield{author}{\bibinfo{person}{Mark Brown}, \bibinfo{person}{Nigel Pope},
  {and} \bibinfo{person}{Kevin Voges}.} \bibinfo{year}{2003}\natexlab{}.
\newblock \showarticletitle{Buying or Browsing? {{An}} Exploration of Shopping
  Orientations and Online Purchase Intention}.
\newblock \bibinfo{journal}{\emph{European Journal of Marketing}}
  \bibinfo{volume}{37}, \bibinfo{number}{11/12} (\bibinfo{date}{Dec.}
  \bibinfo{year}{2003}).
\newblock


\bibitem[\protect\citeauthoryear{Brown and Levinson}{Brown and
  Levinson}{1987}]%
        {brown_politeness_1987}
\bibfield{author}{\bibinfo{person}{Penelope Brown} {and}
  \bibinfo{person}{Stephen~C Levinson}.} \bibinfo{year}{1987}\natexlab{}.
\newblock \bibinfo{booktitle}{\emph{Politeness: Some Universals in Language
  Usage}}.
\newblock \bibinfo{publisher}{{Cambridge University Press}}.
\newblock


\bibitem[\protect\citeauthoryear{Cadilhac, Asher, Benamara, and
  Lascarides}{Cadilhac et~al\mbox{.}}{2013}]%
        {cadilhac_grounding_2013}
\bibfield{author}{\bibinfo{person}{Anais Cadilhac}, \bibinfo{person}{Nicholas
  Asher}, \bibinfo{person}{Farah Benamara}, {and} \bibinfo{person}{Alex
  Lascarides}.} \bibinfo{year}{2013}\natexlab{}.
\newblock \showarticletitle{Grounding {{Strategic Conversation}}: {{Using
  Negotiation Dialogues}} to {{Predict Trades}} in a {{Win}}-{{Lose Game}}}. In
  \bibinfo{booktitle}{\emph{Proceedings of {{EMNLP}}}}.
\newblock


\bibitem[\protect\citeauthoryear{Carvalho}{Carvalho}{2011}]%
        {carvalho_modeling_2011}
\bibfield{author}{\bibinfo{person}{Vitor~R. Carvalho}.}
  \bibinfo{year}{2011}\natexlab{}.
\newblock \showarticletitle{Modeling {{Intention}} in {{Email}} - {{Speech
  Acts}}, {{Information Leaks}} and {{Recommendation Models}}}.
\newblock In \bibinfo{booktitle}{\emph{Studies in {{Computational
  Intelligence}}}}.
\newblock


\bibitem[\protect\citeauthoryear{Chang and {Danescu-Niculescu-Mizil}}{Chang and
  {Danescu-Niculescu-Mizil}}{2019}]%
        {chang_trajectories_2019}
\bibfield{author}{\bibinfo{person}{Jonathan~P. Chang} {and}
  \bibinfo{person}{Cristian {Danescu-Niculescu-Mizil}}.}
  \bibinfo{year}{2019}\natexlab{}.
\newblock \showarticletitle{Trajectories of {{Blocked Community Members}}:
  {{Redemption}}, {{Recidivism}} and {{Departure}}}. In
  \bibinfo{booktitle}{\emph{Proceedings of {{WWW}}}}.
\newblock


\bibitem[\protect\citeauthoryear{Clark}{Clark}{1996}]%
        {clark_using_1996}
\bibfield{author}{\bibinfo{person}{Herbert~H Clark}.}
  \bibinfo{year}{1996}\natexlab{}.
\newblock \bibinfo{booktitle}{\emph{Using Language} (\bibinfo{edition}{second}
  ed.)}.
\newblock \bibinfo{publisher}{{Cambridge University Press}}.
\newblock


\bibitem[\protect\citeauthoryear{Collobert, Weston, Bottou, Karlen,
  Kavukcuoglu, and Kuksa}{Collobert et~al\mbox{.}}{2011}]%
        {collobert_natural_2011}
\bibfield{author}{\bibinfo{person}{Ronan Collobert}, \bibinfo{person}{Jason
  Weston}, \bibinfo{person}{L{\'e}on Bottou}, \bibinfo{person}{Michael Karlen},
  \bibinfo{person}{Koray Kavukcuoglu}, {and} \bibinfo{person}{Pavel Kuksa}.}
  \bibinfo{year}{2011}\natexlab{}.
\newblock \showarticletitle{Natural {{Language Processing}} ({{Almost}}) from
  {{Scratch}}}.
\newblock \bibinfo{journal}{\emph{Journal of Machine Learning Research}}
  \bibinfo{volume}{12}, \bibinfo{number}{Aug} (\bibinfo{year}{2011}).
\newblock


\bibitem[\protect\citeauthoryear{{Corral-Verdugo}}{{Corral-Verdugo}}{1993}]%
        {corral-verdugo_effect_1993}
\bibfield{author}{\bibinfo{person}{Victor {Corral-Verdugo}}.}
  \bibinfo{year}{1993}\natexlab{}.
\newblock \showarticletitle{The {{Effect}} of {{Examples}} and {{Gender}} on
  {{Third Graders}}' {{Ability}} to {{Distinguish Environmental Facts}} from
  {{Opinions}}}.
\newblock \bibinfo{journal}{\emph{The Journal of Environmental Education}}
  \bibinfo{volume}{24}, \bibinfo{number}{4} (\bibinfo{date}{July}
  \bibinfo{year}{1993}).
\newblock


\bibitem[\protect\citeauthoryear{Craker and March}{Craker and March}{2016}]%
        {craker_dark_2016}
\bibfield{author}{\bibinfo{person}{Naomi Craker} {and} \bibinfo{person}{Evita
  March}.} \bibinfo{year}{2016}\natexlab{}.
\newblock \showarticletitle{The Dark Side of {{Facebook}}\textregistered{}:
  {{The Dark Tetrad}}, Negative Social Potency, and Trolling Behaviours}.
\newblock \bibinfo{journal}{\emph{Personality and Individual Differences}}
  \bibinfo{volume}{102} (\bibinfo{date}{Nov.} \bibinfo{year}{2016}).
\newblock


\bibitem[\protect\citeauthoryear{Curhan and Pentland}{Curhan and
  Pentland}{2007}]%
        {curhan_thin_2007}
\bibfield{author}{\bibinfo{person}{Jared~R. Curhan} {and} \bibinfo{person}{Alex
  Pentland}.} \bibinfo{year}{2007}\natexlab{}.
\newblock \showarticletitle{Thin {{Slices}} of {{Negotiation}}: {{Predicting
  Outcomes From Conversational Dynamics Within}} the {{First}} 5 {{Minutes}}.}
\newblock \bibinfo{journal}{\emph{Journal of Applied Psychology}}
  \bibinfo{volume}{92} (\bibinfo{date}{May} \bibinfo{year}{2007}).
\newblock


\bibitem[\protect\citeauthoryear{Devlin, Chang, Lee, and Toutanova}{Devlin
  et~al\mbox{.}}{2019}]%
        {devlin_bert_2019}
\bibfield{author}{\bibinfo{person}{Jacob Devlin}, \bibinfo{person}{Ming-Wei
  Chang}, \bibinfo{person}{Kenton Lee}, {and} \bibinfo{person}{Kristina
  Toutanova}.} \bibinfo{year}{2019}\natexlab{}.
\newblock \showarticletitle{{{BERT}}: {{Pre}}-Training of {{Deep Bidirectional
  Transformers}} for {{Language Understanding}}}. In
  \bibinfo{booktitle}{\emph{Proceedings of {{NAACL}}}}.
\newblock


\bibitem[\protect\citeauthoryear{Filieri}{Filieri}{2016}]%
        {filieri_what_2016}
\bibfield{author}{\bibinfo{person}{Raffaele Filieri}.}
  \bibinfo{year}{2016}\natexlab{}.
\newblock \showarticletitle{What Makes an Online Consumer Review Trustworthy?}
\newblock \bibinfo{journal}{\emph{Annals of Tourism Research}}
  \bibinfo{volume}{58} (\bibinfo{date}{May} \bibinfo{year}{2016}).
\newblock


\bibitem[\protect\citeauthoryear{Gao and Huang}{Gao and Huang}{2017}]%
        {gao_detecting_2017}
\bibfield{author}{\bibinfo{person}{Lei Gao} {and} \bibinfo{person}{Ruihong
  Huang}.} \bibinfo{year}{2017}\natexlab{}.
\newblock \showarticletitle{Detecting {{Online Hate Speech Using Context Aware
  Models}}}. In \bibinfo{booktitle}{\emph{Proceedings of {{RANLP}}}}.
\newblock


\bibitem[\protect\citeauthoryear{Garimella, De~Francisci~Morales, Gionis, and
  Mathioudakis}{Garimella et~al\mbox{.}}{2017}]%
        {garimella_quantifying_2017}
\bibfield{author}{\bibinfo{person}{Kiran Garimella}, \bibinfo{person}{Gianmarco
  De~Francisci~Morales}, \bibinfo{person}{Aristides Gionis}, {and}
  \bibinfo{person}{Michael Mathioudakis}.} \bibinfo{year}{2017}\natexlab{}.
\newblock \showarticletitle{Quantifying {{Controversy}} in {{Social Media}}}.
\newblock \bibinfo{journal}{\emph{ACM Transactions on Social Computing}}
  \bibinfo{volume}{1}, \bibinfo{number}{1} (\bibinfo{year}{2017}).
\newblock


\bibitem[\protect\citeauthoryear{Ghosh, Richard~Fabbri, and Muresan}{Ghosh
  et~al\mbox{.}}{2017}]%
        {ghosh_role_2017}
\bibfield{author}{\bibinfo{person}{Debanjan Ghosh}, \bibinfo{person}{Alexander
  Richard~Fabbri}, {and} \bibinfo{person}{Smaranda Muresan}.}
  \bibinfo{year}{2017}\natexlab{}.
\newblock \showarticletitle{The {{Role}} of {{Conversation Context}} for
  {{Sarcasm Detection}} in {{Online Interactions}}}. In
  \bibinfo{booktitle}{\emph{Proceedings of {{SIGDIAL}}}}.
\newblock


\bibitem[\protect\citeauthoryear{Giambra, Camp, and Grodsky}{Giambra
  et~al\mbox{.}}{1992}]%
        {giambra_curiosity_1992}
\bibfield{author}{\bibinfo{person}{Leonard~M. Giambra},
  \bibinfo{person}{Cameron~J. Camp}, {and} \bibinfo{person}{Alicia Grodsky}.}
  \bibinfo{year}{1992}\natexlab{}.
\newblock \showarticletitle{Curiosity and Stimulation Seeking across the Adult
  Life Span: {{Cross}}-Sectional and 6- to 8-Year Longitudinal Findings}.
\newblock \bibinfo{journal}{\emph{Psychology and Aging}} \bibinfo{volume}{7},
  \bibinfo{number}{1} (\bibinfo{year}{1992}).
\newblock


\bibitem[\protect\citeauthoryear{{Gonz{\'a}lez-Ib{\'a}{\~n}ez}, Muresan, and
  Wacholder}{{Gonz{\'a}lez-Ib{\'a}{\~n}ez} et~al\mbox{.}}{2011}]%
        {gonzalez-ibanez_identifying_2011}
\bibfield{author}{\bibinfo{person}{Roberto {Gonz{\'a}lez-Ib{\'a}{\~n}ez}},
  \bibinfo{person}{Smaranda Muresan}, {and} \bibinfo{person}{Nina Wacholder}.}
  \bibinfo{year}{2011}\natexlab{}.
\newblock \showarticletitle{Identifying {{Sarcasm}} in {{Twitter}}: {{A Closer
  Look}}}. In \bibinfo{booktitle}{\emph{Proceedings of {{ACL}}}}.
\newblock


\bibitem[\protect\citeauthoryear{Grosz and Sidner}{Grosz and Sidner}{1986}]%
        {grosz_attention_1986}
\bibfield{author}{\bibinfo{person}{Barbara~J. Grosz} {and}
  \bibinfo{person}{Candace~L. Sidner}.} \bibinfo{year}{1986}\natexlab{}.
\newblock \showarticletitle{Attention, {{Intentions}}, and the {{Structure}} of
  {{Discourse}}}.
\newblock \bibinfo{journal}{\emph{Computational Linguistics}}
  \bibinfo{volume}{12}, \bibinfo{number}{3} (\bibinfo{date}{July}
  \bibinfo{year}{1986}).
\newblock


\bibitem[\protect\citeauthoryear{Guess, Nagler, and Tucker}{Guess
  et~al\mbox{.}}{2019}]%
        {guess_less_2019}
\bibfield{author}{\bibinfo{person}{Andrew Guess}, \bibinfo{person}{Jonathan
  Nagler}, {and} \bibinfo{person}{Joshua Tucker}.}
  \bibinfo{year}{2019}\natexlab{}.
\newblock \showarticletitle{Less than You Think: {{Prevalence}} and Predictors
  of Fake News Dissemination on {{Facebook}}}.
\newblock \bibinfo{journal}{\emph{Science Advances}} \bibinfo{volume}{5},
  \bibinfo{number}{1} (\bibinfo{date}{Jan.} \bibinfo{year}{2019}).
\newblock


\bibitem[\protect\citeauthoryear{Guy, Makarenkov, Hazon, Rokach, and
  Shapira}{Guy et~al\mbox{.}}{2018}]%
        {guy_identifying_2018}
\bibfield{author}{\bibinfo{person}{Ido Guy}, \bibinfo{person}{Victor
  Makarenkov}, \bibinfo{person}{Niva Hazon}, \bibinfo{person}{Lior Rokach},
  {and} \bibinfo{person}{Bracha Shapira}.} \bibinfo{year}{2018}\natexlab{}.
\newblock \showarticletitle{Identifying {{Informational}} vs. {{Conversational
  Questions}} on {{Community Question Answering Archives}}}. In
  \bibinfo{booktitle}{\emph{Proceedings of {{WSDM}}}}.
\newblock


\bibitem[\protect\citeauthoryear{Hanjalic, Kofler, and Larson}{Hanjalic
  et~al\mbox{.}}{2012}]%
        {hanjalic_intent_2012}
\bibfield{author}{\bibinfo{person}{Alan Hanjalic}, \bibinfo{person}{Christoph
  Kofler}, {and} \bibinfo{person}{Martha Larson}.}
  \bibinfo{year}{2012}\natexlab{}.
\newblock \showarticletitle{Intent and Its Discontents: The User at the Wheel
  of the Online Video Search Engine}. In \bibinfo{booktitle}{\emph{Proceedings
  of {{MM}}}}.
\newblock


\bibitem[\protect\citeauthoryear{Harper, Moy, and Konstan}{Harper
  et~al\mbox{.}}{2009}]%
        {harper_facts_2009}
\bibfield{author}{\bibinfo{person}{F.~Maxwell Harper}, \bibinfo{person}{Daniel
  Moy}, {and} \bibinfo{person}{Joseph~A. Konstan}.}
  \bibinfo{year}{2009}\natexlab{}.
\newblock \showarticletitle{Facts or Friends?: Distinguishing Informational and
  Conversational Questions in Social {{Q}}\&{{A}} Sites}. In
  \bibinfo{booktitle}{\emph{Proceedings of {{CHI}}}}.
\newblock


\bibitem[\protect\citeauthoryear{Hashemi, Asiaee, and Kraft}{Hashemi
  et~al\mbox{.}}{2016}]%
        {hashemi_query_2016}
\bibfield{author}{\bibinfo{person}{Homa~B Hashemi}, \bibinfo{person}{Amir
  Asiaee}, {and} \bibinfo{person}{Reiner Kraft}.}
  \bibinfo{year}{2016}\natexlab{}.
\newblock \showarticletitle{Query {{Intent Detection}} Using {{Convolutional
  Neural Networks}}}. In \bibinfo{booktitle}{\emph{Proceedings of {{QRUMS}}}}.
\newblock


\bibitem[\protect\citeauthoryear{Hessel and Lee}{Hessel and Lee}{2019}]%
        {hessel_somethings_2019}
\bibfield{author}{\bibinfo{person}{Jack Hessel} {and} \bibinfo{person}{Lillian
  Lee}.} \bibinfo{year}{2019}\natexlab{}.
\newblock \showarticletitle{Something's {{Brewing}}! {{Early Prediction}} of
  {{Controversy}}-Causing {{Posts}} from {{Discussion Features}}}. In
  \bibinfo{booktitle}{\emph{Proceedings of {{NAACL}}}}.
\newblock


\bibitem[\protect\citeauthoryear{Jhaver, Appling, Gilbert, and Bruckman}{Jhaver
  et~al\mbox{.}}{2019}]%
        {jhaver_did_2019}
\bibfield{author}{\bibinfo{person}{Shagun Jhaver},
  \bibinfo{person}{Darren~Scott Appling}, \bibinfo{person}{Eric Gilbert}, {and}
  \bibinfo{person}{Amy Bruckman}.} \bibinfo{year}{2019}\natexlab{}.
\newblock \showarticletitle{``{{Did You Suspect}} the {{Post Would}} Be
  {{Removed}}?'': {{Understanding User Reactions}} to {{Content Removals}} on
  {{Reddit}}}. In \bibinfo{booktitle}{\emph{Proceedings of {{CSCW}}}},
  Vol.~\bibinfo{volume}{1}.
\newblock


\bibitem[\protect\citeauthoryear{Jin and Martin}{Jin and Martin}{2015}]%
        {jin_match_2015}
\bibfield{author}{\bibinfo{person}{Seunga~Venus Jin} {and}
  \bibinfo{person}{Cassie Martin}.} \bibinfo{year}{2015}\natexlab{}.
\newblock \showarticletitle{``{{A Match Made}}\ldots{{Online}}?'' {{The
  Effects}} of {{User}}-{{Generated Online Dater Profile Types}}
  ({{Free}}-{{Spirited Versus Uptight}}) on {{Other Users}}' {{Perception}} of
  {{Trustworthiness}}, {{Interpersonal Attraction}}, and {{Personality}}}.
\newblock \bibinfo{journal}{\emph{Cyberpsychology, Behavior, and Social
  Networking}} \bibinfo{volume}{18}, \bibinfo{number}{6} (\bibinfo{date}{June}
  \bibinfo{year}{2015}).
\newblock


\bibitem[\protect\citeauthoryear{Kehler and Rohde}{Kehler and Rohde}{2017}]%
        {kehler_evaluating_2017}
\bibfield{author}{\bibinfo{person}{Andrew Kehler} {and} \bibinfo{person}{Hannah
  Rohde}.} \bibinfo{year}{2017}\natexlab{}.
\newblock \showarticletitle{Evaluating an {{Expectation}}-{{Driven
  Question}}-{{Under}}-{{Discussion Model}} of {{Discourse Interpretation}}}.
\newblock \bibinfo{journal}{\emph{Discourse Processes}} \bibinfo{volume}{54},
  \bibinfo{number}{3} (\bibinfo{date}{April} \bibinfo{year}{2017}).
\newblock


\bibitem[\protect\citeauthoryear{Kim}{Kim}{2014}]%
        {kim_convolutional_2014}
\bibfield{author}{\bibinfo{person}{Yoon Kim}.} \bibinfo{year}{2014}\natexlab{}.
\newblock \showarticletitle{Convolutional {{Neural Networks}} for {{Sentence
  Classification}}}. In \bibinfo{booktitle}{\emph{Proceedings of {{EMNLP}}}}.
\newblock


\bibitem[\protect\citeauthoryear{Kumar, Mahdian, and McGlohon}{Kumar
  et~al\mbox{.}}{2010}]%
        {kumar_dynamics_2010}
\bibfield{author}{\bibinfo{person}{Ravi Kumar}, \bibinfo{person}{Mohammad
  Mahdian}, {and} \bibinfo{person}{Mary McGlohon}.}
  \bibinfo{year}{2010}\natexlab{}.
\newblock \showarticletitle{Dynamics of Conversations}. In
  \bibinfo{booktitle}{\emph{Proceedings of {{KDD}}}}.
\newblock


\bibitem[\protect\citeauthoryear{Lex, Juffinger, and Granitzer}{Lex
  et~al\mbox{.}}{2010}]%
        {lex_objectivity_2010}
\bibfield{author}{\bibinfo{person}{Elisabeth Lex}, \bibinfo{person}{Andreas
  Juffinger}, {and} \bibinfo{person}{Michael Granitzer}.}
  \bibinfo{year}{2010}\natexlab{}.
\newblock \showarticletitle{Objectivity Classification in Online Media}. In
  \bibinfo{booktitle}{\emph{Proceedings of {{HT}}}}.
\newblock


\bibitem[\protect\citeauthoryear{Lex, Voelske, Errecalde, Ferretti, Cagnina,
  Horn, Stein, and Granitzer}{Lex et~al\mbox{.}}{2012}]%
        {lex_measuring_2012}
\bibfield{author}{\bibinfo{person}{Elisabeth Lex}, \bibinfo{person}{Michael
  Voelske}, \bibinfo{person}{Marcelo Errecalde}, \bibinfo{person}{Edgardo
  Ferretti}, \bibinfo{person}{Leticia Cagnina}, \bibinfo{person}{Christopher
  Horn}, \bibinfo{person}{Benno Stein}, {and} \bibinfo{person}{Michael
  Granitzer}.} \bibinfo{year}{2012}\natexlab{}.
\newblock \showarticletitle{Measuring the Quality of Web Content Using Factual
  Information}. In \bibinfo{booktitle}{\emph{Proceedings of {{WebQuality}}}}.
\newblock


\bibitem[\protect\citeauthoryear{Lin, He, and Everson}{Lin
  et~al\mbox{.}}{2011}]%
        {lin_sentence_2011}
\bibfield{author}{\bibinfo{person}{Chenghua Lin}, \bibinfo{person}{Yulan He},
  {and} \bibinfo{person}{Richard Everson}.} \bibinfo{year}{2011}\natexlab{}.
\newblock \showarticletitle{Sentence {{Subjectivity Detection}} with
  {{Weakly}}-{{Supervised Learning}}}. In \bibinfo{booktitle}{\emph{Proceedings
  of {{IJCNLP}}}}.
\newblock


\bibitem[\protect\citeauthoryear{Lin, Kang, Gamon, and Pantel}{Lin
  et~al\mbox{.}}{2018}]%
        {lin_actionable_2018}
\bibfield{author}{\bibinfo{person}{Chu-Cheng Lin}, \bibinfo{person}{Dongyeop
  Kang}, \bibinfo{person}{Michael Gamon}, {and} \bibinfo{person}{Patrick
  Pantel}.} \bibinfo{year}{2018}\natexlab{}.
\newblock \showarticletitle{Actionable {{Email Intent Modeling With
  Reparametrized RNNs}}}. In \bibinfo{booktitle}{\emph{Proceedings of
  {{AAAI}}}}.
\newblock


\bibitem[\protect\citeauthoryear{Ling, Chai, and Piew}{Ling
  et~al\mbox{.}}{2010}]%
        {ling_effects_2010}
\bibfield{author}{\bibinfo{person}{Kwek~Choon Ling}, \bibinfo{person}{Lau~Teck
  Chai}, {and} \bibinfo{person}{Tan~Hoi Piew}.}
  \bibinfo{year}{2010}\natexlab{}.
\newblock \showarticletitle{The {{Effects}} of {{Shopping Orientations}},
  {{Online Trust}} and {{Prior Online Purchase Experience}} toward
  {{Customers}}' {{Online Purchase Intention}}}.
\newblock \bibinfo{journal}{\emph{International Business Research}}
  \bibinfo{volume}{3}, \bibinfo{number}{3} (\bibinfo{date}{June}
  \bibinfo{year}{2010}).
\newblock


\bibitem[\protect\citeauthoryear{Liu}{Liu}{2010}]%
        {liu_sentiment_2010}
\bibfield{author}{\bibinfo{person}{Bing Liu}.} \bibinfo{year}{2010}\natexlab{}.
\newblock \showarticletitle{Sentiment {{Analysis}} and {{Subjectivity}}}.
\newblock In \bibinfo{booktitle}{\emph{Handbook of {{Natural Language
  Processing}}} (\bibinfo{edition}{2nd} ed.)}.
\newblock


\bibitem[\protect\citeauthoryear{Liu, Guberman, Hemphill, and Culotta}{Liu
  et~al\mbox{.}}{2018}]%
        {liu_forecasting_2018}
\bibfield{author}{\bibinfo{person}{Ping Liu}, \bibinfo{person}{Joshua
  Guberman}, \bibinfo{person}{Libby Hemphill}, {and} \bibinfo{person}{Aron
  Culotta}.} \bibinfo{year}{2018}\natexlab{}.
\newblock \showarticletitle{Forecasting the {{Presence}} and {{Intensity}} of
  {{Hostility}} on {{Instagram Using Linguistic}} and {{Social Features}}}. In
  \bibinfo{booktitle}{\emph{Proceedings of {{ICWSM}}}}.
\newblock


\bibitem[\protect\citeauthoryear{Liu and Jansen}{Liu and Jansen}{2015}]%
        {liu_taxonomy_2015}
\bibfield{author}{\bibinfo{person}{Zhe Liu} {and} \bibinfo{person}{Bernard~J.
  Jansen}.} \bibinfo{year}{2015}\natexlab{}.
\newblock \showarticletitle{A {{Taxonomy}} for {{Classifying Questions Asked}}
  in {{Social Question}} and {{Answering}}}. In
  \bibinfo{booktitle}{\emph{Proceedings of {{CHI Extended Abstracts}}}}.
\newblock


\bibitem[\protect\citeauthoryear{McKee}{McKee}{2002}]%
        {mckee_your_2002}
\bibfield{author}{\bibinfo{person}{Heidi McKee}.}
  \bibinfo{year}{2002}\natexlab{}.
\newblock \showarticletitle{``{{YOUR VIEWS SHOWED TRUE IGNORANCE}}!!!'':
  ({{Mis}}){{Communication}} in an Online Interracial Discussion Forum}.
\newblock \bibinfo{journal}{\emph{Computers and Composition}}
  \bibinfo{volume}{19}, \bibinfo{number}{4} (\bibinfo{date}{Dec.}
  \bibinfo{year}{2002}).
\newblock


\bibitem[\protect\citeauthoryear{Mitchell, Gottfried, Barthel, and
  Sumida}{Mitchell et~al\mbox{.}}{2018}]%
        {mitchell_distinguishing_2018}
\bibfield{author}{\bibinfo{person}{Amy Mitchell}, \bibinfo{person}{Jeffrey
  Gottfried}, \bibinfo{person}{Michael Barthel}, {and} \bibinfo{person}{Nami
  Sumida}.} \bibinfo{year}{2018}\natexlab{}.
\newblock \bibinfo{title}{Distinguishing {{Between Factual}} and {{Opinion
  Statements}} in the {{News}}}.
\newblock
  \bibinfo{howpublished}{https://www.journalism.org/2018/06/18/distinguishing-between-factual-and-opinion-statements-in-the-news/}.
\newblock


\bibitem[\protect\citeauthoryear{Moor, Heuvelman, and Verleur}{Moor
  et~al\mbox{.}}{2010}]%
        {moor_flaming_2010}
\bibfield{author}{\bibinfo{person}{Peter~J. Moor}, \bibinfo{person}{Ard
  Heuvelman}, {and} \bibinfo{person}{Ria Verleur}.}
  \bibinfo{year}{2010}\natexlab{}.
\newblock \showarticletitle{Flaming on {{YouTube}}}.
\newblock \bibinfo{journal}{\emph{Computers in Human Behavior}}
  \bibinfo{volume}{26}, \bibinfo{number}{6} (\bibinfo{date}{Nov.}
  \bibinfo{year}{2010}).
\newblock


\bibitem[\protect\citeauthoryear{Morris, Teevan, and Panovich}{Morris
  et~al\mbox{.}}{2010}]%
        {morris_what_2010}
\bibfield{author}{\bibinfo{person}{Meredith~Ringel Morris},
  \bibinfo{person}{Jaime Teevan}, {and} \bibinfo{person}{Katrina Panovich}.}
  \bibinfo{year}{2010}\natexlab{}.
\newblock \showarticletitle{What Do People Ask Their Social Networks, and Why?:
  A Survey Study of Status Message Q\&a Behavior}. In
  \bibinfo{booktitle}{\emph{Proceedings of {{CHI}}}}.
\newblock


\bibitem[\protect\citeauthoryear{Murray and Carenini}{Murray and
  Carenini}{2011}]%
        {murray_subjectivity_2011}
\bibfield{author}{\bibinfo{person}{Gabriel Murray} {and}
  \bibinfo{person}{Giuseppe Carenini}.} \bibinfo{year}{2011}\natexlab{}.
\newblock \showarticletitle{Subjectivity Detection in Spoken and Written
  Conversations}.
\newblock \bibinfo{journal}{\emph{Natural Language Engineering}}
  \bibinfo{volume}{17}, \bibinfo{number}{3} (\bibinfo{date}{July}
  \bibinfo{year}{2011}).
\newblock


\bibitem[\protect\citeauthoryear{Quarteroni, Ivanov, and Riccardi}{Quarteroni
  et~al\mbox{.}}{2011}]%
        {quarteroni_simultaneous_2011}
\bibfield{author}{\bibinfo{person}{Silvia Quarteroni},
  \bibinfo{person}{Alexei~V. Ivanov}, {and} \bibinfo{person}{Giuseppe
  Riccardi}.} \bibinfo{year}{2011}\natexlab{}.
\newblock \showarticletitle{Simultaneous Dialog Act Segmentation and
  Classification from Human-Human Spoken Conversations}. In
  \bibinfo{booktitle}{\emph{Proceedings of {{ICASSP}}}}.
\newblock


\bibitem[\protect\citeauthoryear{Quirk, Greenbaum, Leech, and Svartvik}{Quirk
  et~al\mbox{.}}{1985}]%
        {quirk_comprehensive_1985}
\bibfield{author}{\bibinfo{person}{Randolph Quirk}, \bibinfo{person}{Sidney
  Greenbaum}, \bibinfo{person}{Geoffrey Leech}, {and} \bibinfo{person}{Jan
  Svartvik}.} \bibinfo{year}{1985}\natexlab{}.
\newblock \bibinfo{booktitle}{\emph{A {{Comprehensive Grammar}} of the
  {{English Language}}}}.
\newblock \bibinfo{publisher}{{Longman}}, \bibinfo{address}{{New York}}.
\newblock


\bibitem[\protect\citeauthoryear{Rabinowitz, Acevedo, Casen, Rosengarten,
  Kowalczyk, and Portnoy}{Rabinowitz et~al\mbox{.}}{2013}]%
        {rabinowitz_distinguishing_2013}
\bibfield{author}{\bibinfo{person}{Mitchell Rabinowitz}, \bibinfo{person}{Maria
  Acevedo}, \bibinfo{person}{Sara Casen}, \bibinfo{person}{Myriah Rosengarten},
  \bibinfo{person}{Martha Kowalczyk}, {and} \bibinfo{person}{Lindsay~Blau
  Portnoy}.} \bibinfo{year}{2013}\natexlab{}.
\newblock \showarticletitle{Distinguishing Facts from Beliefs: Fuzzy
  Categories}.
\newblock \bibinfo{journal}{\emph{Psychology of Language and Communication}}
  \bibinfo{volume}{17}, \bibinfo{number}{3} (\bibinfo{date}{Dec.}
  \bibinfo{year}{2013}).
\newblock


\bibitem[\protect\citeauthoryear{Regmi and Bal}{Regmi and Bal}{2015}]%
        {regmi_what_2015}
\bibfield{author}{\bibinfo{person}{Santosh Regmi} {and}
  \bibinfo{person}{Bal~Krishna Bal}.} \bibinfo{year}{2015}\natexlab{}.
\newblock \showarticletitle{What {{Make Facts Stand Out}} from {{Opinions}}?
  {{Distinguishing Facts}} from {{Opinions}} in {{News Media}}}.
\newblock \bibinfo{journal}{\emph{Creativity in Intelligent Technologies and
  Data Science}}  \bibinfo{volume}{535} (\bibinfo{year}{2015}).
\newblock


\bibitem[\protect\citeauthoryear{Reyes, Rosso, and Buscaldi}{Reyes
  et~al\mbox{.}}{2012}]%
        {reyes_humor_2012}
\bibfield{author}{\bibinfo{person}{Antonio Reyes}, \bibinfo{person}{Paolo
  Rosso}, {and} \bibinfo{person}{Davide Buscaldi}.}
  \bibinfo{year}{2012}\natexlab{}.
\newblock \showarticletitle{From Humor Recognition to Irony Detection: {{The}}
  Figurative Language of Social Media}.
\newblock \bibinfo{journal}{\emph{Data \& Knowledge Engineering}}
  \bibinfo{volume}{74} (\bibinfo{date}{April} \bibinfo{year}{2012}).
\newblock


\bibitem[\protect\citeauthoryear{Riloff, Wiebe, and Phillips}{Riloff
  et~al\mbox{.}}{2005}]%
        {riloff_exploiting_2005}
\bibfield{author}{\bibinfo{person}{Ellen Riloff}, \bibinfo{person}{Janyce
  Wiebe}, {and} \bibinfo{person}{William Phillips}.}
  \bibinfo{year}{2005}\natexlab{}.
\newblock \showarticletitle{Exploiting {{Subjectivity Classification}} to
  {{Improve Information Extraction}}}. In \bibinfo{booktitle}{\emph{Proceedings
  of {{AAAI}}}}.
\newblock


\bibitem[\protect\citeauthoryear{Sundar}{Sundar}{1998}]%
        {sundar_effect_1998}
\bibfield{author}{\bibinfo{person}{S.~Shyam Sundar}.}
  \bibinfo{year}{1998}\natexlab{}.
\newblock \showarticletitle{Effect of {{Source Attribution}} on {{Perception}}
  of {{Online News Stories}}}.
\newblock \bibinfo{journal}{\emph{Journalism \& Mass Communication Quarterly}}
  \bibinfo{volume}{75}, \bibinfo{number}{1} (\bibinfo{date}{March}
  \bibinfo{year}{1998}).
\newblock


\bibitem[\protect\citeauthoryear{Tannen}{Tannen}{2000}]%
        {tannen_indirectness_2000}
\bibfield{author}{\bibinfo{person}{Deborah Tannen}.}
  \bibinfo{year}{2000}\natexlab{}.
\newblock \showarticletitle{Indirectness at {{Work}}}.
\newblock In \bibinfo{booktitle}{\emph{Language in {{Action}}: {{New Studies}}
  of {{Language}} in {{Society}}, {{Festschrift}} for {{Roger Shuy}}}}.
\newblock


\bibitem[\protect\citeauthoryear{Tannen}{Tannen}{2005}]%
        {tannen_conversational_2005}
\bibfield{author}{\bibinfo{person}{Deborah Tannen}.}
  \bibinfo{year}{2005}\natexlab{}.
\newblock \bibinfo{booktitle}{\emph{Conversational Style : Analyzing Talk among
  Friends}}.
\newblock \bibinfo{publisher}{{Oxford University Press}},
  \bibinfo{address}{{New York}}.
\newblock


\bibitem[\protect\citeauthoryear{Valliant}{Valliant}{1993}]%
        {valliant_poststratification_1993}
\bibfield{author}{\bibinfo{person}{Richard Valliant}.}
  \bibinfo{year}{1993}\natexlab{}.
\newblock \showarticletitle{Poststratification and {{Conditional Variance
  Estimation}}}.
\newblock \bibinfo{journal}{\emph{J. Amer. Statist. Assoc.}}
  \bibinfo{volume}{88}, \bibinfo{number}{421} (\bibinfo{date}{March}
  \bibinfo{year}{1993}).
\newblock


\bibitem[\protect\citeauthoryear{van~der Heijden, Verhagen, and
  Creemers}{van~der Heijden et~al\mbox{.}}{2001}]%
        {heijden_predicting_2001}
\bibfield{author}{\bibinfo{person}{Hans van~der Heijden},
  \bibinfo{person}{Tibert Verhagen}, {and} \bibinfo{person}{Marcel Creemers}.}
  \bibinfo{year}{2001}\natexlab{}.
\newblock \showarticletitle{Predicting Online Purchase Behavior: Replications
  and Tests of Competing Models}. In \bibinfo{booktitle}{\emph{Proceedings of
  {{HICSS}}}}.
\newblock


\bibitem[\protect\citeauthoryear{Wang, Hosseini, Awadallah, Bennett, and
  Quirk}{Wang et~al\mbox{.}}{2019}]%
        {wang_context-aware_2019}
\bibfield{author}{\bibinfo{person}{Wei Wang}, \bibinfo{person}{Saghar
  Hosseini}, \bibinfo{person}{Ahmed~Hassan Awadallah}, \bibinfo{person}{Paul~N.
  Bennett}, {and} \bibinfo{person}{Chris Quirk}.}
  \bibinfo{year}{2019}\natexlab{}.
\newblock \showarticletitle{Context-{{Aware Intent Identification}} in {{Email
  Conversations}}}. In \bibinfo{booktitle}{\emph{Proceedings of {{SIGIR}}}}.
\newblock


\bibitem[\protect\citeauthoryear{Wiebe, Breck, Buckley, Cardie, Davis, Fraser,
  Litman, Pierce, Riloff, Wilson, Day, and Maybury}{Wiebe
  et~al\mbox{.}}{2003}]%
        {wiebe_recognizing_2003}
\bibfield{author}{\bibinfo{person}{Janyce Wiebe}, \bibinfo{person}{Eric Breck},
  \bibinfo{person}{Chris Buckley}, \bibinfo{person}{Claire Cardie},
  \bibinfo{person}{Paul Davis}, \bibinfo{person}{Bruce Fraser},
  \bibinfo{person}{Diane Litman}, \bibinfo{person}{David Pierce},
  \bibinfo{person}{Ellen Riloff}, \bibinfo{person}{Theresa Wilson},
  \bibinfo{person}{David Day}, {and} \bibinfo{person}{Mark Maybury}.}
  \bibinfo{year}{2003}\natexlab{}.
\newblock \showarticletitle{Recognizing and {{Organizing Opinions Expressed}}
  in the {{World Press}}}. In \bibinfo{booktitle}{\emph{{{AAAI Symposium}} on
  {{New Directions}} in {{Question Answering}}}}.
\newblock


\bibitem[\protect\citeauthoryear{Wiebe, Wilson, Bruce, Bell, and Martin}{Wiebe
  et~al\mbox{.}}{2004}]%
        {wiebe_learning_2004}
\bibfield{author}{\bibinfo{person}{Janyce Wiebe}, \bibinfo{person}{Theresa
  Wilson}, \bibinfo{person}{Rebecca Bruce}, \bibinfo{person}{Matthew Bell},
  {and} \bibinfo{person}{Melanie Martin}.} \bibinfo{year}{2004}\natexlab{}.
\newblock \showarticletitle{Learning {{Subjective Language}}}.
\newblock \bibinfo{journal}{\emph{Computational Linguistics}}
  \bibinfo{volume}{30}, \bibinfo{number}{3} (\bibinfo{date}{Sept.}
  \bibinfo{year}{2004}).
\newblock


\bibitem[\protect\citeauthoryear{Wiebe, Wilson, and Cardie}{Wiebe
  et~al\mbox{.}}{2005}]%
        {wiebe_annotating_2005}
\bibfield{author}{\bibinfo{person}{Janyce Wiebe}, \bibinfo{person}{Theresa
  Wilson}, {and} \bibinfo{person}{Claire Cardie}.}
  \bibinfo{year}{2005}\natexlab{}.
\newblock \showarticletitle{Annotating {{Expressions}} of {{Opinions}} and
  {{Emotions}} in {{Language}}}.
\newblock \bibinfo{journal}{\emph{Language Resources and Evaluation}}
  \bibinfo{volume}{39}, \bibinfo{number}{2} (\bibinfo{date}{May}
  \bibinfo{year}{2005}).
\newblock


\bibitem[\protect\citeauthoryear{Wiebe, Bruce, and O'Hara}{Wiebe
  et~al\mbox{.}}{1999}]%
        {wiebe_development_1999}
\bibfield{author}{\bibinfo{person}{Janyce~M. Wiebe},
  \bibinfo{person}{Rebecca~F. Bruce}, {and} \bibinfo{person}{Thomas~P.
  O'Hara}.} \bibinfo{year}{1999}\natexlab{}.
\newblock \showarticletitle{Development and Use of a Gold-Standard Data Set for
  Subjectivity Classifications}. In \bibinfo{booktitle}{\emph{Proceedings of
  {{ACL}}}}.
\newblock


\bibitem[\protect\citeauthoryear{Yessenalina, Yue, and Cardie}{Yessenalina
  et~al\mbox{.}}{2010}]%
        {yessenalina_multi-level_2010}
\bibfield{author}{\bibinfo{person}{Ainur Yessenalina}, \bibinfo{person}{Yisong
  Yue}, {and} \bibinfo{person}{Claire Cardie}.}
  \bibinfo{year}{2010}\natexlab{}.
\newblock \showarticletitle{Multi-Level {{Structured Models}} for
  {{Document}}-Level {{Sentiment Classification}}}. In
  \bibinfo{booktitle}{\emph{Proceedings of {{EMNLP}}}}.
  \bibinfo{address}{{Cambridge, Massachusetts}}.
\newblock


\bibitem[\protect\citeauthoryear{Yu and Hatzivassiloglou}{Yu and
  Hatzivassiloglou}{2003}]%
        {yu_towards_2003}
\bibfield{author}{\bibinfo{person}{Hong Yu} {and} \bibinfo{person}{Vasileios
  Hatzivassiloglou}.} \bibinfo{year}{2003}\natexlab{}.
\newblock \showarticletitle{Towards {{Answering Opinion Questions}}:
  {{Separating Facts}} from {{Opinions}} and {{Identifying}} the {{Polarity}}
  of {{Opinion Sentences}}}. In \bibinfo{booktitle}{\emph{Proceedings of
  {{EMNLP}}}}.
\newblock


\bibitem[\protect\citeauthoryear{Zhang, Chang, {Danescu-Niculescu-Mizil},
  Dixon, Thain, Hua, and Taraborelli}{Zhang et~al\mbox{.}}{2018}]%
        {zhang_conversations_2018}
\bibfield{author}{\bibinfo{person}{Justine Zhang}, \bibinfo{person}{Jonathan~P.
  Chang}, \bibinfo{person}{Cristian {Danescu-Niculescu-Mizil}},
  \bibinfo{person}{Lucas Dixon}, \bibinfo{person}{Nithum Thain},
  \bibinfo{person}{Yiqing Hua}, {and} \bibinfo{person}{Dario Taraborelli}.}
  \bibinfo{year}{2018}\natexlab{}.
\newblock \showarticletitle{Conversations {{Gone Awry}}: {{Detecting Early
  Signs}} of {{Conversational Failure}}}. In
  \bibinfo{booktitle}{\emph{Proceedings of {{ACL}}}}.
\newblock


\bibitem[\protect\citeauthoryear{Zhang, Spirling, and
  {Danescu-Niculescu-Mizil}}{Zhang et~al\mbox{.}}{2017}]%
        {zhang_asking_2017}
\bibfield{author}{\bibinfo{person}{Justine Zhang}, \bibinfo{person}{Arthur
  Spirling}, {and} \bibinfo{person}{Cristian {Danescu-Niculescu-Mizil}}.}
  \bibinfo{year}{2017}\natexlab{}.
\newblock \showarticletitle{Asking Too {{Much}}? {{The Rhetorical Role}} of
  {{Questions}} in {{Political Discourse}}}. In
  \bibinfo{booktitle}{\emph{Proceedings of {{EMNLP}}}}.
\newblock


\bibitem[\protect\citeauthoryear{Zhang and LeCun}{Zhang and LeCun}{2015}]%
        {zhang_text_2015}
\bibfield{author}{\bibinfo{person}{Xiang Zhang} {and} \bibinfo{person}{Yann
  LeCun}.} \bibinfo{year}{2015}\natexlab{}.
\newblock \bibinfo{booktitle}{\emph{Text {{Understanding}} from {{Scratch}}}}.
\newblock \bibinfo{type}{{T}echnical {R}eport}.
\newblock


\end{thebibliography}
\end{document}